\documentclass{ws-ijmpe}
\usepackage[super,compress]{cite}
\usepackage{epstopdf}
\renewcommand\a{\alpha}
\renewcommand\b{\beta}
\renewcommand\d{\delta}

\renewcommand\o{\omega}

\newcommand\g{\gamma}

\newcommand\m{\mu}
\newcommand\n{\nu}
\newcommand\p{\pi}

\renewcommand\L{\Lambda}
\renewcommand\P{\Pi}
\renewcommand\S{\Sigma}

\newcommand\D{\Delta}
\newcommand\G{\Gamma}

\newcommand{\fig}[1]{Fig.~\ref{#1}}
\newcommand{\eq}[1]{Eq.~(\ref{#1})}

\newcommand\lb{\left(}
\newcommand\rb{\right)}
\newcommand\ls{\left[}
\newcommand\rs{\right]}

\newcommand{\lan}{\langle}
\newcommand{\ran}{\rangle}

\newcommand\ra{\rightarrow}

\newcommand{\non}{\nonumber\\}
\newcommand\pt{\partial}

\newcommand{\diag}{{\rm{diag}}}

\newcommand{\im}{{\rm{Im}}}
\newcommand{\bx}{{\mathbf x}}

\newcommand{\bp}{{\mathbf p}}
\newcommand{\bk}{{\mathbf k}}
\newcommand{\bq}{{\mathbf q}}

\newcommand{\bv}{{\bf v}}

\newcommand{\bB}{{\bf B}}
\newcommand{\bE}{{\bf E}}

\begin{document}

\markboth{Authors' Names}{Instructions for
Typing Manuscripts (Paper's Title)}

\catchline{}{}{}{}{}

\title{KINETIC EVOLUTION OF THE GLASMA AND  \\ THERMALIZATION IN HEAVY ION COLLISIONS
}

\author{\footnotesize XU-GUANG HUANG
}

\address{Physics Department and Center for Particle Physics and Field Theory, Fudan University, Shanghai 200433, China.
\\
huangxuguang@fudan.edu.cn}

\author{JINFENG LIAO
}

\address{Physics Department and Center for Exploration of Energy and Matter,
Indiana University,\\
2401 N Milo B. Sampson Lane, Bloomington, IN 47408, USA.\\
RIKEN BNL Research Center, Bldg. 510A, Brookhaven National Laboratory,\\
   Upton, NY 11973, USA. \\
liaoji@indiana.edu}

\maketitle

\begin{history}
\received{Day Month Year}
\revised{Day Month Year}
\end{history}

\begin{abstract}
In relativistic heavy ion collisions, a highly occupied gluonic matter is created shortly after initial impact, which is in a non-thermal state and often referred to as the Glasma. Successful phenomenology suggests that the glasma evolves rather quickly toward the thermal quark-gluon plasma and a hydrodynamic behavior emerges at very early time $\sim \hat{o}(1) \, \rm fm/c$. Exactly how such ``apparent thermalization'' occurs and connects the initial conditions to the hydrodynamic onset, remains a significant challenge for theory as well as phenomenology. We briefly review various ideas and recent progress in understanding the approach of the glasma to the thermalized quark-gluon plasma, with an emphasis on the kinetic theory description for the evolution of such far-from-equilibrium and highly overpopulated, thus weakly-coupled yet strongly interacting glasma.
\end{abstract}


\ccode{PACS numbers:12.38.Mh, 25.75.-q, 05.60.-k, 03.75.Nt}


\section{Introduction}

Relativistic heavy ion collisions provide the unique way for creating and measuring new forms of strongly interacting matter. Such experiments are now carried out at both the Relativistic Heavy Ion Collider (RHIC)~\cite{Adams:2005dq,Arsene:2004fa,Back:2004je,Adcox:2004mh} and the Large Hadron Collider (LHC)~\cite{Muller:2012zq}. In such collisions, two large nuclei (e.g. gold or lead ions) are accelerated to move at nearly the speed of light and collide with each other, creating a domain of matter with extremely high energy density well exceeding the expected energy density for the transition from confined hadronic matter to deconfined strongly interacting quark-gluon matter. This matter subsequently evolves toward a thermal quark-gluon plasma (QGP)~\cite{Gyulassy:2004zy,Shuryak:2004cy} experiencing hydrodynamic expansion~\cite{Heinz:2009xj,Heinz:2013th,Huovinen:2013wma,Romatschke:2009im}. The QGP expands in a viscous-hydrodynamic way and eventually cools down enough to hadronize into the hadronic gas that further expands and ultimately freezes out into thousands of individual hadrons measured by detectors. The evolution during such collisions is highly dynamical and involves the thermal, near-thermal (transport), as well as far-from-thermal properties of the created strongly interacting matter.

The focus of this brief review is the so-called ``thermalization'' problem~\cite{Berges:2012ks,Gelis:2012ri,Strickland:2013uga,Arnold:2007pg}, which is about how the system evolves from the initial condition to the nearly thermal QGP, in a relatively short time at the order of $\sim \hat{o}(1)\, \rm fm/c$. In order to understand the context, let us first discuss what is before and after this short transient period. Before collision, the initial state of the large nuclei at very high energy can be relatively well understood and described by the so-called color glass condensate (CGC) effective theory, with an inherent momentum scale called the saturation scale $Q_s$. The scale $Q_s$ grows with collisional beam energy and with the size of nuclei, and for RHIC and LHC collisions the relevant scale is on the order of a few $\rm GeV$, which implies that the relevant QCD coupling $\alpha_s$ is not large and weak coupling description should be feasible at least initially. On the other hand, successful phenomenology based on hydrodynamic simulations of the fireball evolution have provided accurate and detailed descriptions of an incredible amount of data from RHIC to LHC. Essentially all such simulations require the hydrodynamic stage to start at a very short time, typically $0.5\sim 1 \, \rm fm/c$, after the initial collision. The onset of hydrodynamic behavior is usually assumed as an indication of nearly local thermal equilibration, so the system seems to have a very short pre-equilibrium evolution stage in between the initial state and the hydrodynamic onset. Normally a short relaxation time $\tau \sim 1/(\alpha_s^2Q_s)$ toward equilibrium would indicate large coupling, which apparently is in tension with the relatively high scale $Q_s$ thus small coupling from the initial state. To make it even more complicated, there is strong longitudinal expansion from the very beginning of the evolution which constantly drives system toward significant anisotropy between longitudinal and transverse pressures: to what degree the system maintains (an)isotropy and how, remain as open questions. To date, a precise understanding of evolution toward the (apparent and approximate) thermalization in such pre-equilibrium matter, called the ``glasma'' (in between the color glass condensate and the thermal quark-gluon plasma), is still lacking.  The thermalization problem thus presents both an outstanding theoretical challenge and a significant phenomenological gap.

With more than a decade's study on this problem, many different ideas and approaches have been proposed and developed and varied mechanisms are found to play certain roles. These include e.g. the kinetic evolutions emphasizing either elastic or inelastic or both processes, the various plasma instabilities, the real time lattice simulations based on classical statistical field theories, and more recently the strongly coupled scenarios based on gauge/gravity duality framework. It is difficult to sufficiently discuss all these in the present paper, and there already exist excellent sources covering one or more aspects of these. For recent reviews, see e.g. \cite{Berges:2012ks,Gelis:2012ri,Strickland:2013uga,Arnold:2007pg}. Instead, we will focus on the discussions of the transport approach, with an emphasis on some recent nontrivial results in the kinetic evolution of the far-from-equilibrium and highly overpopulated glasma.

The rest of the paper is organized as follows. In Section 2, we will briefly survey the general context and the key issues in the pre-equilibrium evolution, including discussions on approaches other than the kinetic one. In Section 3, the kinetic framework for describing the glasma will be introduced and different scattering processes  will be discussed. The Section 4 will summarize some recent results on the kinetic evolution of the glasma that make the nontrivial link from the initial overpopulation to possible dynamical, transient, Bose-Einstein Condensation. The Section 5 will discuss results from other kinetic approaches. Finally the summary and some concluding remarks will be given in Section 6.

\section{Pre-Equilibrium Evolution in Heavy Ion Collisions}

The pre-equilibrium evolution in a heavy ion collision is complicated and may involve a few stages, from the CGC~\cite{McLerran:1993ni,McLerran:1993ka,Gribov:1984tu,Mueller:1985wy,Blaizot:1987nc,Iancu:2003xm,Iancu:2002xk,Weigert:2005us,Gelis:2010nm}, through an anisotropic strong field stage  \cite{Lappi:2006fp,Gelis:2009wh}, toward initial isotropization via instabilities of various kinds \cite{Weibel:1959zz,Mrowczynski:1988dz,Mrowczynski:1993qm,Mrowczynski:2000ed,Mrowczynski:2005ki,Arnold:2003rq,Arnold:2004ti,Arnold:2004ih,Arnold:2007cg,Romatschke:2003ms,Romatschke:2004jh,Romatschke:2005ag,Romatschke:2005pm,Romatschke:2006wg,Romatschke:2006nk,Rebhan:2004ur,Rebhan:2005re,Rebhan:2008uj,Rebhan:2009ku,Bodeker:2005nv,Bodeker:2007fw,Dumitru:2006pz,Berges:2008zt,Berges:2012cj,Gelis:2013rba,Epelbaum:2013waa,Dusling:2012ig} till the time scale $\sim 1/Q_{\rm s}$ (up to factors logarithmic in coupling). These plasma instabilities are ``triggered'' by the very initial anisotropy and the rapid growth of unstable models frees up the quanta from classical fields and redistributes momentum among different directions thus bringing the system back to be near isotropy. From thereon the system could also be considered as a dense system of gluons that becomes amenable to  kinetic evolution toward local equilibration  based upon  the Boltzmann transport approach ~\cite{Mueller:1999fp,Mueller:1999pi,Baier:2000sb,Baier:2002qv,Mueller:2005un,Mueller:2005hj,Mueller:2006up,Arnold:2003rq,Xu:2004mz,Blaizot:2011xf,Blaizot:2013lga,Huang:2013lia,Kurkela:2011ti,Kurkela:2011ub}. In this section, we will briefly discuss these different stages, some of the key issues, and various approaches (other than the kinetic one which will be the focus of the next two sections).

\subsection{Pre-collision: the Color Glass Condensate}

Unlike the Big Bang, the one-shot start of our Universe, for which we can not know what preceded it, for the heavy ion collisions known as the Little Bang, we have a good understanding of the high energy nuclei coming into the collisions and we can have such collisions repeatedly in laboratories. As it turns out, the gluonic content of a nucleon or nucleus at very high energy enters the so-called saturation regime, described by the Color Glass Condensate effective theory (see reviews in e.g. ~\cite{Gelis:2012ri,Iancu:2003xm,Iancu:2002xk,Weigert:2005us,Gelis:2010nm}). To see how the saturation arises, consider the gluon distribution of a nucleon moving at extremely high energy $E=\sqrt{s}/2$ with respect to the lab frame. The extreme high energy brings in significant Lorentz dilation effect (with $\gamma=E/M$, $M$ the nucleon mass) on all the intrinsic time scales of the nucleon, in particular 1) the lifetime of a quantum fluctuation $\delta\tau \sim 1/\delta E \to \gamma \delta\tau$ from which all sea gluons (and quarks) originate, and 2) the time scale of interactions among all these valence and sea partons. Therefore when viewed at very high energy, the nucleon looks like a very dense system of nearly free ``wee'' partons. In addition, going to very high energy allows probing partons that carry very small fraction of longitudinal momentum of the nucleon $x=p_z / E \to 0$, i.e. the small-$x$ region of parton distributions. Measurements from Deep Inelastic Scatterings (e.g. at HERA~\cite{Aaron:2009aa}) have shown indeed that there is a very rapid growth of gluon numbers in the small-$x$ region that overwhelmingly dominates over all other parton species. So there is a very dense system of gluons emerging at small $x$ inside the high energy nucleon/nucleus.

The growth of gluon numbers at small-$x$, however, can not continue forever. At high enough density of these gluons, the recombination starts to become important and ultimately brings such growth to stop at a maximal density $\sim 1/\alpha_s$, i.e. saturation. Suppose the gluon density (on the transverse plane of the highly contracted nucleon/nucleus) probed at given $x$ and transverse resolution scale $Q$ is $xG(x,Q^2)$ then an intrinsic {\it saturation scale $Q_s$} emerges and the onset of saturation, $xG / Q^2 \to  1/\alpha_s  $ happens for all scales $Q\le Q_s$, with
\begin{eqnarray}
Q_s^2 \equiv \ \alpha_s\,  xG(x,Q_s^2)
\end{eqnarray}
This saturation scale changes with the nuclear size, $Q_s^2 \sim A^{1/3}$, as well as with $x$, $Q_s^2 \sim 1/x^{-0.3}$. Therefore for very large nucleus colliding at very high energy, the $Q_s$ becomes very large $Q_s\gg \Lambda_{QCD}$ thus allowing a weak-coupling based description of the saturated dense gluon system. Estimates suggest that $Q_s\sim 1-2 \rm GeV$ for RHIC AuAu collisions and $Q_s\sim 2-3 \rm GeV$ for LHC PbPb collisions.

In the CGC description based on the McLerran-Venugopalan model~\cite{McLerran:1993ni,McLerran:1993ka}, one separates the fast and slow partons in the fast-moving nucleon with certain cutoff scale in longitudinal momentum. By virtue of Lorentz dilation the fast partons can be treated as approximately independent color sources with certain color charge distribution $\rho^a$ and only subject to local correlations. The slow partons (dominantly gluons) as a dense saturated system with the phase space density $f\sim xG/Q_s^2 \sim 1/\alpha_s$ can then be treated as classical fields from solving Yang-Mills field equations with the presence of such color charge distribution. The cutoff dependence is governed by proper evolution equations. Of course, quantum fluctuations dictate that such source distribution $\rho^a$ differs in each collision event. So one needs to specify a whole ensemble of the charge density distribution based on certain probability distribution $W[\rho]$ (e.g. Guassian) together with the classical field configurations solved for each specific $\rho$. With this machinery, proper initial conditions from the colliding nuclei for the heavy ion collisions can be provided.

Let us just emphasize two important features of the pre-collision color glass condensate: first, the emergence of an intrinsic scale, $Q_s$; second, the saturated phase space density $f\sim 1/\alpha_s$ for gluons seen at $Q<Q_s$. One may naturally imagine these two features being inherited by the very initial stage of the glasma, which is true albeit through a rather indirect way as we discuss next.

\subsection{The initial Glasma fields, instability, and isotropization}

With the initial states of colliding nuclei described by the GCC framework, let us then examine the system in the collision zone just after collision ($\tau=0^+$). This can be done by numerically solving the Yang-Mills equations in the forward light-cone with the given sources in that collision, i.e.
\begin{eqnarray}
[D_\mu , F^{\mu\nu}] = J^\nu
\end{eqnarray}
with $J^\nu$ given by the fast moving color charge densities $\rho_{1,2}(x_\perp)$ on the two light cones from the two initial nuclei.
A striking finding is that the classical color fields are basically electric and magnetic fields in parallel to the collision beam axis (in $\hat{z}$ direction) with vanishing transverse components, i.e. $\bE^a = E^a \hat{z}$ and $\bB^a = B^a \hat{z}$, just like color flux tubes stretching between (random) color sources in the two sheets of nuclei moving apart from the collision point. The corresponding stress tensor associated with such field configurations takes the form $T^{\mu\nu} = \diag(\epsilon,\epsilon,\epsilon,-\epsilon)$, i.e. with {\it negative longitudinal pressure} that is obviously far from a hydrodynamic form. So, the initial glasma fields are highly anisotropic. Such anisotropy however does not last for long, due to the {\it instabilities}~\cite{Mrowczynski:1993qm,Mrowczynski:2000ed,Mrowczynski:2005ki,Romatschke:2005pm,Romatschke:2005ag,Romatschke:2006wg,Romatschke:2006nk,Dumitru:2006pz,Rebhan:2005re}.

The various plasma instabilities generically arise as a consequence of anisotropy and leads to exponential growth of modes that help restore the isotropy. In a sense, just like particle scatterings in a gas always tend to randomize and thus isotropize the momentum distribution, the classical fields have interactions built in and it is not surprising that the fluctuations on top of the fields ``know'' which direction to involve toward. The interesting feature, however, is the {\it exponential} behavior which is significantly more efficient than usual scattering processes. This has been quantitatively studied in several approaches, such as the semi-classical transport in the hard-loop framework~\cite{Rebhan:2005re,Rebhan:2008uj} or the classical-statistical lattice simulations of the field evolution~\cite{Berges:2008zt,Berges:2012cj,Gelis:2013rba,Epelbaum:2013waa}. For simplicity let us take the classical-statistical field theory approach as the example here. On top of the purely longitudinal boost-invariant initial fields, one may introduce rapidity-dependent quantum fluctuations that evolve in the background initial fields. The solutions exhibit exponential growth of such fluctuations for characteristic modes (with $(\#)\sim \hat{o}(1)$ constant)
\begin{eqnarray}
\delta A \sim e^{(\#)\, \sqrt{(\#) \, Q_s \tau}}
\end{eqnarray}
This sets a limiting time scale at which the {\it quantum fluctuations become as large as the initial background classic fields}, i.e. $\delta A (\tau_s)\to A_0 \sim 1/g$: $\tau_s \sim (1/Q_s) \, \ln^2(1/g)$. The evolution from the initial collision till this limiting time scale for instabilities is of course rather complicated, but is in principle computable with ab initio first-principle approach at sufficiently weak coupling. For the purpose of our discussions, let us just mention the following important features regarding the system at the time scale $\tau_s$: 1) as the result of (primary and secondary) instabilities, a wide range of modes, up to the order of saturation momentum, grow until reaching the saturated regime with non-perturbatively large occupation number $\sim 1/\alpha_s$ --- this in a sense inherits the characteristics of the saturated initial gluon distribution in an indirect way; 2) the growth of these modes builds up the longitudinal pressure and isotropizes the stress tensor $T^{\mu\nu} = \diag(\epsilon,P_T,P_T,P_L)$ to the extent  of possible remaining $\hat o(1)$ anisotropy between $P_L$ and $P_T$.

\subsection{Field evolution from classical-statistical lattice simulations}

From this point on, i.e. $\tau>\tau_s\sim (1/Q_s) \, \ln^2(1/g)$, the system becomes a very dense system of fluctuation  modes (in the statistical-field language) or equivalently gluons (in the kinetic language) with high occupation $\sim 1/\alpha_s$. The initial high anisotropy in glasma fields has by now been reduced to be rather mild. Studies on the further evolution of such a system toward equilibration may be divided into two categories. One category deals with the system's evolution in a fixed volume i.e. without expansion (which is often referred to as ``static box case''). This is certainly of great theoretical interest and in fact quite challenging. For example, a final conclusion is yet to be achieved even regarding the seemingly simple parametric question of thermalization time $\tau_{\rm th} \sim \alpha_s^{(\#?)}/Q_s$ in the static box case. The studies of static box case also provide extremely useful insights on the roles of various driving mechanisms toward thermalization. The other category deals with the evolution of the system undergoing boost-invariant longitudinal expansion (often referred to as ``expanding case''), which is a more realistic setting relevant to heavy ion collisions.

Both the static box case and the expanding case have been thoroughly studied using the classical-statistical lattice simulations, for the scalar field theories as well as the Yang-Mills theories: see e.g. most recent results in~\cite{Berges:2012cj,Gelis:2013rba,Epelbaum:2011pc,Gelis:2011xw,Berges:2012us,Berges:2011sb,Berges:2012ev,Berges:2012iw,Kurkela:2012hp,Schlichting:2012es,Berges:2013eia,Berges:2013fga,Berges:2013lsa,York:2014wja} . Within this approach's regime of validity i.e. weak coupling $\a_s\ll1$ and high occupation $f\gg1$, these studies have provided fairly detailed pictures of the field evolution, with the stress tensor components and spectrum of their correlators also evaluated. Among other interesting results, it was found that the evolution is characterized by a pair of dual cascades: the particle cascade toward the infrared modes, and the energy cascade toward the ultraviolet modes. On the ultraviolet end, the simulations running toward very large time (which implies very small coupling to ensure the validity of this approach at late time) appears to show the system's evolution onto non-thermal fixed point with a self-similar form for which the scaling exponents could be understood via turbulent scaling arguments~\cite{Berges:2013eia,Berges:2013fga,Berges:2013lsa}. On the infrared end, simulations (for the scalar field case) starting with overpopulated initial conditions appear to show clear evidences~\cite{Epelbaum:2011pc,Berges:2012us} for the onset of a dynamically formed out-of-equilibrium Bose-Einstein Condensate as predicted in ~\cite{Blaizot:2011xf,Blaizot:2013lga} .

Let us discuss a little bit about the complications in the expanding case. In the static box case, there is a well-defined thermal fixed point to which the system will eventually equilibrate, and conservation laws (e.g. for energy, and for particle number in purely elastic case) are straightforward. When the system undergoes boost-invariant longitudinal expansion, the situation is quite different as strictly speaking there is no well-defined static thermal fixed point. Both the energy and the particle number are dropping with time due to expansion that dilutes the system.  What's more the expansion is constantly bringing the system out of isotropy, i.e. even if the system starts in an isotropic state it will quickly become anisotropic due to ``shrinking of longitudinal momenta''. Note this is a dynamical issue quite separated and different from the high anisotropy from glasma fields in the very initial condition. If there is no interaction, then the system will continuously free-stream with less and less average longitudinal momentum compared with the transverse one, $\lan p^2_L\ran \,  \ll \, \lan p^2_T\ran$. Of course interactions, e.g. scatterings, help re-distribute the momentum between the longitudinal and the transverse and try to restore isotropy to some extent. The issue is, whether the scatterings are strong enough to compete with the longitudinal expansion. It could be that the expansion wins and the system anisotropy constantly grows till falling apart~\cite{Baier:2000sb,Baier:2002qv}. It could also be that the scatterings are able to keep the anisotropy below or at most of order one for a long time~\cite{Blaizot:2011xf} . To complicate it further, how the energy density decreases with time depends on the degree of anisotropy, while on the other hand the energy density dictates the hard scale that would affect the efficiency of scatterings against expansion: so this is a highly dynamical, and nonlinear issue. Classical-statistical lattice simulations at extremely small coupling (e.g. $\alpha_s \sim 10^{-5}$) appear to suggest the scenario of growing anisotropy~\cite{Berges:2013eia,Berges:2013fga,Berges:2013lsa}. However while moving to the relatively larger coupling regime (e.g. $\alpha_s \sim 10^{-2}$) but still within the applicability of the approach, there appears to be a plausible transition to a different behavior of the evolution in which the pressure anisotropy is maintained at order one and able to be matched to viscous hydrodynamics~\cite{Gelis:2013rba} . At the moment a final conclusion is yet to be reached, and in particular a lot of future efforts will be required to push the classical field approach toward the physically more relevant regime (with $\alpha_s \sim 10^{-1}$) which is a highly nontrivial challenge. As a final comment, it seems quite clear that with the presence of longitudinal expansion, a full isotropization may never have been reached, and an anisotropy at least of order one may be present for a considerable window in the early time dynamics of heavy ion collisions. Such ``fixed anisotropy'' for microscopically long but macroscopically short time scale may provide an underlying basis for the recently developed anisotropic hydrodynamics (aHydro)~\cite{Martinez:2010sc,Martinez:2010sd,Martinez:2012tu,Bazow:2013ifa} framework that explicitly accounts for the  sizable anisotropy at early times.

\subsection{From classical fields to kinetic quanta}

From the discussions above it is clear that the classical field theory description, valid for large occupation number $f\gg1$, will come to an end at some point as  the occupation number $f$ of various modes decreases rapidly with time due to longitudinal expansion. This is of course the process of field decoherence that frees up individual gluons. A natural framework to describe the weakly coupled gluon system is the kinetic theory based on Boltzmann transport equation. In such an approach, one uses a distribution function $f(t, \bx, \bp)$ to effectively describe the system and dynamics (e.g. various scattering processes) enters via collision kernel ${\cal C}[f]$, and with provided an initial condition then it evolves via transport equation ${\cal D}_t f = {\cal C}[f]$ in a definitive manner. The transport approach is good for weak coupling and not too large occupation, $\alpha_s\ll1$ and $f \le  1/\alpha_s$. A complete description of the glasma evolution at weak coupling shall plausibly involve a proper switch from the classical fields to the kinetic quanta. Many works~\cite{Mueller:1999fp,Mueller:1999pi,Baier:2000sb,Baier:2002qv,Mueller:2005un,Mueller:2005hj,Mueller:2006up,Arnold:2003rq,Xu:2004mz,Blaizot:2011xf,Blaizot:2013lga,Huang:2013lia,Kurkela:2011ti,Kurkela:2011ub} have been done to study the pre-equilibrium evolution in the kinetic approach, which will be thoroughly discussed   in the next two Sections.

One may notice that there is an overlapping region for the validity of the classical field versus kinetic approaches: for occupation in the regime $1\ll f \le  1/\alpha_s$, both descriptions are feasible, and therefore should be connected with each other. Indeed, the equivalence of the two has been demonstrated in various theories, see e.g.~\cite{Blaizot:1999xk,Mueller:2002gd,Mueller:2006up,Jeon:1995zm}. Roughly speaking, the equation of motion for the Green's function in classical field theory can be suitably mapped to the evolution equation for distribution function, with the interaction terms in the former becoming the collision terms in the transport equation. This is particularly interesting as one may expect dual descriptions in the two approaches for the same physical phenomenon occurring in their common valid regime. The kinetic approach can oftentimes help develop intuitive pictures for understanding results from classical field simulations. For example, the interesting turbulent scaling exponents from late time non-thermal fixed point found in classical field simulations could be easily understood via the pertinent kinetic description~\cite{Berges:2013eia}. Another example is the occurrence of BEC starting with overpopulated initial condition, which was easier to be predicted first from the kinetic evolution~\cite{Blaizot:2011xf}, while less obvious in the classical field description. It is extremely useful to have such dual descriptions as an important tool to develop deeper understanding and have mutual confirmation for interesting results from each other.

\subsection{Thermalization in strongly-coupled theories}

Finally let us also briefly discuss an ``orthogonal'' approach for understanding thermalization with {\it strong coupling} for the underlying microscopic theory. This is different from what has been discussed so far, where we consider the system to be weakly coupled but strongly interacting due to high phase space density. In strong coupling regime it is difficult to have a direct ``attack'', and instead one utilizes the tool of gauge/gravity duality~\cite{Chesler:2009cy,Wu:2012rib,Wu:2013qi,Chesler:2013lia,Lin:2008rw,Lin:2009pn,Kovchegov:2009du,Beuf:2009cx,Heller:2011ju,Heller:2012je,CaronHuot:2011dr,Chesler:2011ds,Balasubramanian:2013rva,Balasubramanian:2013oga}. This holographic correspondence provides a powerful framework to study the evolution of 4-dimensional strongly coupled quantum fluid toward equilibration subject to varied far-from-equilibrium initial conditions, via solving  5-dimensional classical gravity problems. The time dependence of the 4-D theory is translated into the general relativity dynamics in the 5-D (asymptotically) AdS space with proper boundary conditions. Apart from the issue of to what extent such strongly coupled descriptions are applicable to the early time system in heavy ion collisions, these studies have certainly provided interesting insights into the far-from-equilibrium evolution in quantum field theories.
Not being able to give a detailed discussion here on many interesting results from the holographic approach (which is not the primary focus of the present review), let us emphasize one particularly important point clearly demonstrated in such studies:  by analyzing the evolution of the energy-momentum tensor one sees a viscous-hydrodynamic behavior that emerges quickly, well before the full equilibration, and upon the onset of such hydrodynamic regime the system still bears significant anisotropy between the longitudinal and transverse pressures. That is, an {\it ``apparent macroscopic thermalization''} may occur long preceding the true and complete microscopic thermalization.

\section{Kinetic Description of the Pre-Equilibrium Evolution}

We now turn to an elementary discussion on the kinetic description of the pre-equilibrium evolution, namely, a description based on the transport equation of gluons with various scattering processes. Standard textbooks for kinetic theory include~\cite{lifshitz,degroot,liboff}.

As discussed previously, in heavy ion collisions gluons are freed at a time scale
$t_0\sim Q_s^{-1}$, with momentum typically of order $Q_s$ and  phase space occupation number of order $1/\alpha_s$. For large nuclei and/or high collision energies $Q_s\gg \L_{QCD}$ so that $\alpha_s\ll1$.
After that (i.e., for $Q_st>1$),
one may then treat the gluons as on-shell quanta, and effectively describe the system with a phase space distribution function
\begin{eqnarray}
f(t,\bx,\bp)\equiv\frac{(2\p)^3}{N_g}\frac{dN}{d^3\bx d^3\bp},
\end{eqnarray}
where $N_g=2(N_c^2-1)=16$ is the gluon degeneracy. The kinetic evolution of $f(t,\bx,\bp)$ is described
by the Boltzmann equation which schematically reads
\begin{eqnarray}
\label{boltzmann}
{\cal D}_t f(t,\bx,\bp)={\cal C}[f],
\end{eqnarray}
where
\begin{eqnarray}
{\cal D}_t f(t,\bx,\bp) \equiv\frac{p^\m}{E_p}\pt_\m f(t,\bx,\bp)=(\pt_t+\bv_p\cdot\nabla_\bx) f(t,\bx,\bp)
\end{eqnarray}
with $\bv_p\equiv\bp/E_p$ being the velocity of gluon with momentum $\bp$ and $E_p=|\bp|$ the energy of the gluon. Note that on the left-hand side of the equation we have neglected the force term ${\bf F} \cdot \nabla_{\bf p}$ which need to be included if external color fields may be applied. In the following, unless explicitly stated, we will restrict ourselves to the spatially homogeneous systems so that $f$ has no $\bx$
dependence and we can ignore the drift term $\bv_p\cdot\nabla_\bx f_p$ on the left-hand side of the  equation.

With the general structure of kinetic equation above, there are two important ingredients that govern the solutions to it: (1) the initial condition, i.e. the $f(t_0,\bx,\bp)$ at initial time $t_0$; (2) the dynamics of the underlying microscopic theory, i.e. various collisional processes, that will enter through the collision kernel on the right hand side of the equation. Both play nontrivial roles in the evolution, as we shall see in later sections with more concrete examples.

\subsection{Conservation laws}

The conservation laws play important roles in studies of kinetic evolution. These conservation laws originate from the conservation laws observed by the corresponding microscopic interactions. Let us discuss  a number of such examples.

First of all, let us consider the particle number conservation when there are only elastic scatterings, i.e. the collision term only has contributions from $m\to m$ processes. The number density is given by $n = \int d^3\bp/(2\p)^3 f_p=\int_{\bp}2E_p f_p$ where $f_p\equiv f(t,\bp)$ and
\begin{eqnarray}
\int_\bp&\equiv&\int\frac{d^3\bp}{(2\p)^32E_p}.
\end{eqnarray} Therefore the changing rate of number density is given by (considering all particles being bosons)
\begin{eqnarray}
&& {\cal D}_t n  = \int\frac{d^3\bp_1}{(2\p)^3} {\cal C}[f_1] \propto \int_{{1,2,...,m}} \int_{{m+1,m+2,...,2m}} |M_{m\ra m}|^2 \d^{(4)}\lb \S_{i=1}^m p_i-\S_{j=m+1}^{2m} p_{j}\rb\nonumber \\
&& \quad \times  \{ [\Pi_{i=1}^m(1+f_i)]  [\Pi_{j=m+1}^{2m}f_j]  - [\Pi_{i=1}^m f_i]  [\Pi_{j=m+1}^{2m}(1+f_j)] \} \quad \nonumber \\
&& \quad=0,
\end{eqnarray}
where the right hand side vanishes by symmetry (i.e. with the same number of particles in the initial and final states in the microscopic scatterings). Note also that $ {\cal D}_t n = \partial_\mu \lan n u_\mu\ran=\partial_\mu \int d^3\bp/(2\p)^3f_p u_\mu$ with $u_\mu=p_\mu/E_p$ the four-velocity.

Now let us consider the energy-momentum conservation. The energy momentum tensor is given by $T^{\mu\nu} = \int_{\bp}{p^\mu p^\nu}f_p$ and for generic $m \to n $ processes
\begin{eqnarray}
&& \partial_\mu T^{\mu \nu}=\int\frac{d^3\bp_1}{(2\p)^3} p_1^\n{\cal C}[f_1]
\propto \int_{{1,...,m}} \int_{{m+1,...,m+n}} p_1^\n |M_{m\ra n}|^2\d^{(4)}\lb \S_{i=1}^m  p_i-\S_{j=m+1}^{m+n}p_j\rb \nonumber \\ && \quad\times \{ [\Pi_{i=1}^m(1+f_i)]  [\Pi_{j=m+1}^{m+n}f_j]  - [\Pi_{i=1}^m f_i]  [\Pi_{j=m+1}^{m+n}(1+f_j)] \}\non
&& \quad \propto   \int_{{1,..,m}} \int_{{m+1,...,m+n}} |M_{m\ra n}|^2(\S_{i=1}^m p_i^\nu - \S_{j=m+1}^{m+n}p_j^\nu ) \delta^{(4)}(\S_{i=1}^mp_i - \S_{j=m+1}^{m+n}p_j )\nonumber \\ && \quad\times \{ [\Pi_{i=1}^m(1+f_i)]  [\Pi_{j=m+1}^{m+n}f_j]  - [\Pi_{i=1}^m f_i]  [\Pi_{j=m+1}^{m+n}(1+f_j)] \} \nonumber \\
&& \quad = 0,
\end{eqnarray}
where the right-hand side vanishes by the cyclic symmetry for re-labeling all particles and the microscopic conservation (via the delta function).

If one considers homogeneous systems with the origin at the whole system's center of mass (thus zero total momentum),  then the energy conservation $\partial_t \epsilon = 0$ and particle number conservation $\partial_t n =0$ (in the pure elastic case) are the two important global constraints.

More directly relevant to application toward heavy ion collisions is the situation with boost-invariant longitudinal expansion. In that case, one has to take into account the drift term on the left-hand side (LHS), i.e.
\begin{eqnarray}
{\cal D}_t f(t,\bx,\bp) =(\partial_t + v_z \partial_z)f(t,z,\bp) = {\cal C}[f].
\end{eqnarray}
By assuming boost-invariance $f(t,z,\bp) \to f(\tau, y-\eta, \bp_\perp)$ (where $y$ is momentum rapidity while $\eta$ the spatial rapidity) and by focusing on the system at mid-rapidity $\eta\to 0$ or $z\to 0$, one can simplify the above equation into the following form
\begin{eqnarray} \label{eq_kinetic_boost}
{\cal D}_t f(t,\bx,\bp) =\lb\partial_\tau - \frac{p_z}{\tau} \partial_{p_z}\rb f(\tau,\bp) = \lb\partial_t - \frac{p_z}{t} \partial_{p_z}\rb f(t,\bp) = {\cal C}[f].
\end{eqnarray}
We notice that the above kinetic equation can be rewritten as
\begin{eqnarray}
\lb\partial_t - \frac{p_z}{t} \partial_{p_z}\rb f(t,\bp)  =  \frac{\partial (t\, f)}{t \partial_t} -  \nabla_{\bp} \cdot \left [ \frac{p_z}{t} f  \, \hat{z} \right].
\end{eqnarray}

By integrating the above equation one can easily get the corresponding version for the time evolution of number density $n$ and energy density $\epsilon$:
\begin{eqnarray}
\partial_t n + n /t = 0 \quad \to \quad  n = n_0 \times \frac{t_0}{t}
\end{eqnarray}
\begin{eqnarray}
\partial_t \epsilon + (1+\delta) \epsilon / t = 0 \quad \to \quad  \epsilon = \epsilon_0 \times \left(\frac{ t_0}{t}\right)^{1+\delta}
\end{eqnarray}
where $\delta=P_L/\epsilon$ is the ratio of longitudinal pressure to the energy density that characterizes the system's degree of anisotropy: in isotropic case $\delta=1/3$ and $\epsilon\sim 1/t^{4/3}$ as in ideal hydrodynamics while in free-streaming case $\delta\to 0$ and $\epsilon \to 1/t$.

\subsection{The elastic collision kernel}
The collision kernel ${\cal C}[f]$ controls the rate at which the gluons change their momentum state by collision processes. With gluon scatterings in QCD, the leading order contributions to ${\cal C}[f]$ consist of two essentially different processes, the elastic $2\ra2$ process to be discussed in this subsection,  as well as the  inelastic effective $1\ra2$ process to be discussed in the next subsection.

Let us first look at the $2\ra2$ elastic collision kernel, given by
\begin{eqnarray}
\label{elker}
{\cal C}_{2\ra2}[f_1]&=&\frac{1}{2}\int_{234}\frac{1}{2E_1}|M_{12\ra34}|^2
(2\p)^4\d^{(4)}(p_1+p_2-p_3-p_4)\non&&\times[(1+f_1)(1+f_2)f_3f_4-f_1f_2(1+f_3)(1+f_4)].
\end{eqnarray}
In \eq{elker}, the factor $1/2$ in front of the integral is a symmetry factor, which takes into account the fact that exchanging the gluons 3 and 4 leads to identical configuration being doubly counted in both the matrix element and the full 3 and 4 momentum integration. The $(1+f_i)$ factors represent the final state Bose enhancement arising from quantum nature which is vitally important when $f_i$ becomes large. In the dilute limit $f_i\ll1$ then $(1+f_i) \to 1$ reducing to the classical Boltzmann regime. $|M_{12\ra34}|^2$ is the $2\ra2$ matrix element (in the vacuum),
\begin{eqnarray}
\label{2to2}
|M_{12\ra34}|^2&=&8g^4N_c^2\lb3-\frac{tu}{s^2}-\frac{su}{t^2}-\frac{ts}{u^2}\rb,
\end{eqnarray}
with $s, t, u$ the usual Mandelstam variables
\begin{eqnarray}
s=(p_1+p_2)^2,\;\; t=(p_1-p_3)^2,\;\; u=(p_1-p_4)^2,
\end{eqnarray}
and $p_i=(E_i,\bp_i)$. As is well known for Coulomb type long range interaction, the dominant contribution in the elastic collision integral comes from the small angle scatterings, where there is very small momentum transfer $q\ll p_i$ between the colliding gluons and the incoming states' momenta get deflected only with small angles. This is evident from the matrix element $|M_{12\ra34}|^2$ with divergence in the $u$ and $t$ channels at $u\ra 0$ or $t\ra0$ which ultimately lead to logarithmic contributions. Thus under the small angle approximation, the particle's momentum receives a small but random deflection in each of a series of collisions and experiences a ``random walk'' in the momentum space. Following the Landau-Lifshitz approach~\cite{lifshitz},
the Boltzmann equation with the $2\ra2$ collision kernel can then be reduced to a Fokker-Planck equation
describing such momentum space diffusion, by rewriting ${\cal C}_{2\ra2}[f_1]$ as the divergence of a current ${\bf \cal J}(\bp_1)$ in momentum space,
\begin{eqnarray}
\label{bolel}
\pt_t f_1=-{\bm \nabla}_1\cdot{\bf \cal J}(\bp_1),
\end{eqnarray}
where $\nabla_1=\pt/\pt \bp_1$. A standard calculation yields~\cite{Blaizot:2011xf,Blaizot:2013lga}
\begin{eqnarray}
\label{current}
{\bf\cal J}(\bp)=-\frac{g^4N_c^2L}{8\p^3}\ls I_a{\bm \nabla}f_p+I_b\bv_p f_p(1+f_p)\rs.
\end{eqnarray}
Here $L=\int d|\bq|/|\bq|$ is the Coulomb logarithm which needs to be regularized by the medium screening effect, $L\sim \ln (q_{\rm max}/q_{\rm min})$ where ${q_{\rm max}}$ is typically
of order of the hard scale of the problem (e.g., the temperature if equilibrium is achieved) and $q_{\rm min}$ is determined by the screening mass, $q_{\rm min}\sim m_D$ with $m_D$ the Debye screening mass, $m_D^2\sim\alpha_s \int d^3\bp (f_p/E_p)$.

It shall be noted that the thermal fixed point of Eq.(\ref{bolel}) is precisely the Bose-Einstein distribution  $f_{\rm eq}(\bp)=\frac{1}{\exp{[(E_p-\m)/T}]-1}$. It is also straightforward to show that the derived elastic collision kernel on the right hand side of (\ref{bolel}) conserves both particle number and energy.

The two integrals $I_a$ and $I_b$ are defined as follows
\begin{eqnarray}
\label{intia}
I_a&\equiv&2\p^2\int \frac{d^3\bp}{(2\p)^3}f_p(1+f_p),\\
\label{intib}
I_b&\equiv&2\p^2\int \frac{d^3\bp}{(2\p)^3}\frac{2f_p}{E_p}.
\end{eqnarray}
The integral $I_a$ plays the role of a diffusion constant. For a gluon undergoing  successive random small angle  scatterings over a time window $t$, its momentum will undergo a random walk acquiring a total of final momentum square transfer   $\lan \D p^2 \ran  \sim \hat{q}_{\rm el}t$ where the parameter $\hat{q}_{\rm el}$ characterizes the momentum diffusion. The $\hat{q}_{\rm el}$ is relates to $I_a$ simply by (up to pre-factor in logarithm of $\a_s$)
\begin{eqnarray}
\label{qhatel}
\hat{q}_{\rm el}\sim \a_s^2 I_a.
\end{eqnarray}
The integral $I_b$ is proportional to the Debye screening mass, \begin{eqnarray}
m^2_D \sim \a_sI_b.
\end{eqnarray}
To give concrete examples: in the thermal equilibrium with Bose-Einstein distribution, the two integrals become $I_a \sim T^3$ and $I_b \sim T^2$ and in fact $I_a = T\, I_b$; away from equilibrium this gets changed, e.g. in a glasma-type distribution with $f\sim 1/\a_s$ up to $Q_s$, one gets $I_a \sim Q_s^3/\alpha_s^2$ while $I_b \sim Q_s^2 /\alpha_s$ with different power dependence on coupling.

\subsection{The inelastic collision kernel}\label{kernel23}
We now turn to the inelastic collision kernel. The lowest-order inelastic process of gluon scatterings is the $2\ra3$ process which in naive power counting is $\a_s$ suppressed as compared with the elastic process. This however is rather tricky due to strong infrared divergences present in the corresponding matrix element. In fact, a careful analysis would reveal that its contribution to the collision kernel ${\cal C}[f]$ is at the same parametric order of coupling constant as the elastic process, due to the strong soft and collinear enhancement in the $2\ra3$ matrix element, as will become transparent later. In general, the $2\ra3$ collision kernel takes the following form:
\begin{eqnarray}
{\cal C}_{2\ra3}[f_1]&=&{\cal C}^a_{2\ra3}[f_1]+{\cal C}_{2\ra3}^b[f_1],\\
{\cal C}_{2\ra3}^a[f_1]&=&\frac{1}{6}\int_{2345}\frac{1}{2E_1}|M_{12\ra345}|^2
(2\p)^4\d^4(p_1+p_2-p_3-p_4-p_5)\non&&\times[(1+f_1)(1+f_2)f_3f_4f_5-f_1f_2(1+f_3)(1+f_4)(1+f_5)],\non
{\cal C}_{2\ra3}^b[f_1]&=&\frac{1}{4}\int_{2345}\frac{1}{2E_1}|M_{34\ra125}|^2
(2\p)^4\d^4(p_1+p_2+p_5-p_3-p_4)\non&&\times[(1+f_1)(1+f_2)(1+f_5)f_3f_4-f_1f_2f_5(1+f_3)(1+f_4)].
\end{eqnarray}
The two terms ${\cal C}^a$ and ${\cal C}^b$ differ in that the momentum $p_1$ (that one is ``watching'') is on the two-particle side in the former while on the three-particle side in the latter. The general form of the leading-order $|M_{12\ra345}|^2$ is known~\cite{Berends:1981rb,Ellis:1985er,Gottschalk:1979wq}:
\begin{eqnarray}
\label{M23general}
|M_{12\ra345}|^2&=&g^6N_c^3\frac{\cal N}{\cal D}[(12345)+(12354)+(12435)+(12453)+(12534)+(12543)\non&&+
(13245)+(13254)+(13425)+(13524)+(14235)+(14325)],
\end{eqnarray}
where
\begin{eqnarray}
{\cal N}&=&(12)^4+(13)^4+(14)^4+(15)^4+(23)^4\non&&+(24)^4+(25)^4+(34)^4+(35)^4+(45)^4,\non
{\cal D}&=&(12)(13)(14)(15)(23)(24)(25)(34)(35)(45),\non
(ijklm)&=&(ij)(jk)(kl)(lm)(mi),\non
(ij)&\equiv& p_i\cdot p_j.\nonumber
\end{eqnarray}
Note that $|M_{12\ra345}|^2$ itself is completely symmetric in permutation of $p_i$.

Like the $2\ra2$ matrix element $|M_{12\ra34}|^2$, the $2\ra3$ matrix element $|M_{12\ra345}|^2$ also contains the small angle singularity. In addition, it possesses a more severe IR singularity, the collinear singularity which occurs when the softest gluon of the five particles, say, gluon $5$ moves in a collinear way with one of the other gluons during either absorption or emission. The collinear singularity is also well known in perturbation theory, associated with the massless kinematics of gluons. By collecting the most
singular contributions in $|M_{12\ra345}|^2$, one arrives at the  Gunion-Bertsch formula (e.g. for the piece of $t$-channel and soft $p_5$)~\cite{Gunion:1981qs,Arnold:2000dr,Xu:2007jv,Chen:2009sm,Huang:2013lia,Fochler:2013epa}:
\begin{eqnarray}
\label{2to3}
|M_{12\ra345}|_{\rm GB}^2
&=&16g^6N_c^3\frac{(p_1\cdot p_2)^3}{(p_1\cdot p_3)(p_2\cdot p_4)(p_1\cdot p_5)(p_2\cdot p_5)}.
\end{eqnarray}
When applying the above (specific channel) Gunion-Bertsch formula to ${\cal C}_{2\ra3}^a$, one needs to multiply the kernel by a symmetry factor of  $2\times3=6$ to account for the
other five identical contributions (from $u$-channel and soft $p_3,p_4$ singularities). When
applying it to ${\cal C}_{2\ra3}^b$, there is a symmetry factor of $2\times2=4$ to take into account the other three identical contributions (with one factor $2$ from the degeneracy of $u,t$-channel, another factor $2$ from the degeneracy of soft $p_2, p_5$). As a commonly adopted strategy, one needs to regularize various  IR singularities that survive to the end results of the collision kernel, by appropriate medium screening mass as a IR cutoff in order to obtain finite results.
It should also be noted that the GB approximation can be systematically extended
by including less singular terms order by order through expanding the exact $|M_{12\ra345}|^2$ in
$p_5/\sqrt{s}$ and $t/s$, see Ref.~\cite{Abir:2010kc,Bhattacharyya:2011vy,Fochler:2013epa,Huang:2013lia} for details.

Under the small angle and collinear approximations, the inelastic kernel can be much simplified. Physically there are two types of contributions that can be seen by
examining the softest scale among the external gluons and the internal (exchanging) gluons, say, $q$ and $p_5$. If $p_5\ll q$, the $2\ra3$ process can be regarded as an effective $2\ra2$ process with a slight modification due to final state emission of a very soft gluon $p_5$. On the other hand, if $q\ll p_5$, the $2\ra3$ process can be
considered as an effective $1\ra2$ process with one ``hard'' gluon getting a small ``kick'' and experiencing ``bremsstrahlung''. With this in mind, an analytic inelastic kernel can be derived (see details in \cite{Huang:2013lia}) and the final result  reads:
\begin{eqnarray}
{\cal C}_{2\ra3}[f_1]&=&{\cal C}^{\rm eff}_{2\ra2}[f_1]+{\cal C}_{1\ra2}^{\rm eff}[f_1],
\end{eqnarray}
where
\begin{eqnarray}
{\cal C}_{2\ra2}^{\rm eff}[f_1]&=&\frac{1}{2}\int_{234}\frac{1}{2E_1}|M_{12\ra34}|^2_{\rm eff}
(2\p)^4\d^{(4)}(p_1+p_2-p_3-p_4)\non&&\times[(1+f_1)(1+f_2)f_3f_4-f_1f_2(1+f_3)(1+f_4)],
\end{eqnarray}
and
\begin{eqnarray}
{\cal C}_{1\ra2}^{\rm eff}[f_1]
&=&\int_0^{1}dz|M_{1\ra2}|^2_{\rm eff}
\Big\{\frac{1}{2}[g_{p_1}f_{(1-z){p_1}}f_{z{p_1}}-f_{p_1}g_{(1-z){p_1}}g_{z{p_1}}]\non&&+\frac{1}{z^3}[
f_{p_1/z}g_1g_{(1-z){p_1}/z}-g_{p_1/z}f_{p_1}f_{(1-z){p_1}/z}]\Big\}.
\end{eqnarray}
In the above we have introduced  $$|M_{12\ra34}|^2_{\rm eff}={\cal D}(q)|M_{12\ra34}|^2$$ and $$|M_{1\ra2}|^2_{\rm eff}=\frac{6g^6N_c^3CQI_a}{(2\p)^5}\frac{1}{z(1-z)}$$
where $C=\int_{-1}^1 dx/(1-x)$, $Q=\int dq/q^3$, and
\begin{eqnarray}
\label{dq}
{\cal D}(q)&=&\int_{k<q}\frac{2g^2N_c}{|\bk|^2}
\ls\frac{1+2f_k}{1-\bv_k\cdot\bv_1}+\frac{1+2f_k}{1-\bv_k\cdot\bv_p}\rs.
\end{eqnarray}
Again, the IR divergence in $C$, $Q$ and ${\cal D}$ should be appropriately regularized by screening effects and up to the logarithm of $\a_s$
\begin{eqnarray}
C\sim1,\;\;\; Q\sim 1/m_D^2,
\end{eqnarray}
and ${\cal D}$ is at most $\hat{o}(1)$ order for $f\lesssim1/\a_s$. Thus the main new, number-changing, contribution  of the $2\ra3$ process to the collision kernel, is the effective splitting/joining $1\ra2$ kernel ${\cal C}_{1\ra2}^{\rm eff}$ which can be concisely written as follows by collecting all the order $\hat{o}(1)$ constants into a parameter $R$:
\begin{eqnarray}
\label{eff12}
{\cal C}_{1\ra2}^{\rm eff}[f_1]
&=&R\frac{\a_s^3I_a}{m_D^2}\int_0^{1}dz\frac{1}{z(1-z)}
\Big\{\frac{1}{2}[g_{p_1}f_{(1-z){p_1}}f_{z{p_1}}-f_{p_1}g_{(1-z){p_1}}g_{z{p_1}}]\non&&+\frac{1}{z^3}[
f_{p_1/z}g_1g_{(1-z){p_1}/z}-g_{p_1/z}f_{p_1}f_{(1-z){p_1}/z}]\Big\}.
\end{eqnarray}
It is not difficult to see that: first,  the thermal fixed point of Eq.(\ref{eff12}) is the Bose-Einstein distribution {\it with zero chemical potential}  $f_{\rm eq}(\bp)=1/[\exp{(E_p/T)}-1]$; second, the kernel conserves energy while does not conserve particle number.

Let us now examine the power counting and compare the elastic kernel (\ref{bolel}) versus the inelastic kernel (\ref{eff12}). By noting parametrically the $m_D^2\sim \a_s f^2$ one realizes that both kernels are at the same order. To be more concrete, let us examine both the thermal case and the highly off-equilibrium glasma case. In the thermal case, we have $f\sim 1$, $m_D^2\sim \a_s T^2$, $I_a\sim T^3$ and $I_b\sim T^2$: thus the collision rate for both types of processes are parametrically $\Gamma \sim \a_s^2 T$. In the glasma case we have $f\sim 1/\a_s$ up to $Q_s$, $m_D^2 \sim Q_s^2$, $I_a \sim Q_s^3/\a_s^2$ and $I_b \sim Q_s^2/\a_s$: thus the collision rate for both types of processes are parametrically $\Gamma \sim Q_s$. We therefore see that the two types of processes are indeed contributing to the kinetic evolution at the same parametric order, and hence both need to be included at this order.

\subsection{Higher order kernels and the LPM suppression}\label{lpmeffect}
From last subsection, we have seen that the leading contribution from the  the $2\ra3$ matrix element benefits from soft and collinear enhancement and becomes an  effective
$1\ra2$ process parametrically at the same order in $\a_s$ as the small-angle $2\ra2$ scattering. This contribution,  essentially a bremsstrahlung process, may however bear further complication. Let us consider the emission of a soft gluon with momentum $p_5$ by a parent gluon upon one small ``kick''. The emission needs a formation time $t_{\rm form}(p_5)$ to be completed, though, and during that time it is likely the parent gluon may experience yet another ``kick'': in fact this is inevitable if the formation time $t_{\rm form}(p_5)$ becomes bigger than the ``mean-free-path'' of the parent gluon in the medium, and the emission is ``blended'' together with multiple scatterings. This is of course the well know and well studied Landau-Pomeranchuk-Migdal (LPM) effect~\cite{Landau:1953um,Landau:1953gr,Migdal:1956tc} initially found in QED.  A proper treatment requires resuming the $1+n\ra2+n$ ($n\geq1$) multiple scatterings, and has been understood  in the context of QCD particularly for the jet energy loss~\cite{Baier:1996sk,Baier:1996kr,Zakharov:1996fv,Zakharov:1997uu,Baier:2000mf}.

To take into account the LPM effect in the kinetic equation is a nontrivial task, and a thorough treatment in this aspect has been developed
by Arnold, Moore and Yaffe~\cite{Arnold:2001ba,Arnold:2001ms,Arnold:2002ja,Arnold:2002zm,Arnold:2003zc,Baym:2006qf}. As a schematic approach, one can encode the LPM effect into the Boltzmann equation in the following way. Let the differential splitting/merging rate be $d\G/dk$, then the $1\ra2$ collision kernel in the collinear limit would be
\begin{eqnarray}
\label{general12}
{\cal C}_{1\ra2}[f_p]\sim\int dk\frac{d\G}{dk}
\ls(g_pf_{p-k}f_k-f_pg_{p-k}g_k)
+\lb\frac{p+k}{p}\rb^\eta(g_pg_kf_{p+k}-f_pf_kg_{p+k})\rs,\non
\end{eqnarray}
where $p=|\bp|, k=|\bk|$, and $\bk\parallel\bp$. The parameter $\eta$ will be fixed  by enforcing energy conservation to be preserved by the kernel.  If the formation time $t_{\rm form}$ is shorter than the duration $t_{\rm el}$ between two successive elastic scatterings, the spitting/merging rate is of the Bethe-Heitler type which parametrically reads~\cite{Bethe:1934za}
\begin{eqnarray}
\frac{d\G_{\rm BH}}{dk}\sim\frac{\a_s}{k}\G_{\rm el}\sim\frac{\a_s}{k}\int_{pp'}|M|^2_{2\ra2}f_p(1+f_{p'})\sim\frac{\a_s}{k}\frac{\hat{q}_{\rm el}}{m_D^2},
\end{eqnarray}
where $\G_{\rm el}$ is the rate of soft elastic scattering. Substituting this into \eq{general12}, one arrives at a kernel bearing the structure of \eq{eff12} for a single scattering case. On the other hand, if the formation time $t_{\rm form}$ is longer than $t_{\rm el}$, then the emission process cannot resolve individual collision and ``feels'' the coherent superposition of multiple scatterings during the formation time: in this case, one has $\G_{\rm LPM}(k)\sim \a_s t_{\rm form}^{-1}(k)$. During this formation time, the emitted gluon obtains a transverse momentum $\D k_\perp\sim\sqrt{\hat{q}_{\rm el}t_{\rm form}}$ while its transverse size is $\D l_\perp\sim 1/\D k_\perp$ and its transverse velocity is $v_\perp\sim \D k_\perp/k$. For the emission to be completed, the emitted gluon must separate from its parent gluon, which implies the condition
\begin{eqnarray}
t_{\rm form}\sim\frac{\D l_\perp}{v_\perp}\sim\frac{k}{\hat{q}_{\rm el}t_{\rm form}},
\end{eqnarray}
from which we obtain $t_{\rm form}(k)\sim\sqrt{k/\hat{q}_{\rm el}}$. Thus the LPM suppressed splitting/merging rate would have the form
\begin{eqnarray}
\frac{d\G_{\rm LPM}}{dk}\sim\frac{\a_s}{k}\sqrt{\frac{\hat{q}_{\rm el}}{k}}.
\end{eqnarray}
Substituting it into \eq{general12}, one obtains the following form of the kernel at the end:
\begin{eqnarray}
\label{eff12lpm}
{\cal C}_{1\ra2}^{\rm LPM}[f_p]
&\sim&\a_s^2\sqrt{\frac{I_a}{p}}\Big\{\int_0^{1}\frac{dz}{z^{3/2}}
[g_{p}f_{(1-z){p}}f_{z{p}}-f_{p}g_{(1-z){p}}g_{z{p}}]\non&&+\int_0^{1}\frac{dz}{z^{2}}\frac{1}{[z(1-z)]^{3/2}}[
f_{p/z}g_pg_{(1-z){p}/z}-g_{p/z}f_{p}f_{(1-z){p}/z}]\Big\}.
\end{eqnarray}
Note that the LPM effect plays important role only when $t_{\rm form}(k)\gtrsim t_{\rm el}\sim m_D^2/\hat{q}_{\rm el}$, i.e, when $k$ is larger than $m_D^4/\hat{q}_{\rm el}$. It is clear the above kernel preserves a similar loss/gain structure to the kernel in \eq{eff12}, thus also having the same thermal fixed point. It is also at the same parametric order as (\eq{eff12}) in coupling constant.

\section{Kinetic Evolution in Overpopulated Regime and Possible Bose-Einstein Condensation in the Glasma}

With the kinetic framework set up in the previous Section, we now turn to discuss the application of this framework to the description of the kinetic evolution of the glasma that is pertinent to the early stage in heavy ion collisions. As already discussed previously, since the initial scale $Q_s$ in the glasma is large and thus the coupling is weak, the kinetic theory seems to be a natural and plausible framework to investigate the detailed evolution of the phase space distribution in the dense gluon system starting from the time scale $\sim 1/Q_{\rm s}$. Such efforts were initiated long ago~\cite{Mueller:1999fp,Mueller:1999pi,Baier:2000sb,Baier:2002qv,Mueller:2005un,Mueller:2005hj,Mueller:2006up,Xu:2004mz} and some of these will be discussed in the next Section. An apparent tension in such approaches exists in that in a naive counting the scattering rate (of leading elastic processes) $\sim \alpha_s^2$ may not be able to bring the system back to thermalization quickly enough. A number of past kinetic works suggest that the inelastic processes may play more significant role as compared with the elastic ones in speeding up the thermalization process, especially in populating the very soft momentum region. This may be true in the dilute regime (close to the Boltzmann limit), however may not be the accurate picture when the system under consideration is in the {\it highly overpopulated regime} with $f\sim 1/\alpha_s$.  As shown in a number of recent kinetic studies~\cite{Blaizot:2011xf,Blaizot:2013lga}, the elastic scatterings with highly overpopulated initial conditions can lead to order $\sim {\alpha_s^0}$ evolution and develop strong infrared cascade with the Bose enhancement, and in fact may even induce a dynamical Bose-Einstein Condensation. In this Section we focus on discussing some of these most recent developments.

\subsection{The highly overpopulated Glasma}

To see a few nontrivial features associated with high overpopulation, let us consider the kinetic evolution in a weakly coupled gluon system  initially described by the following glasma-type distribution as inspired by the CGC picture:
\begin{eqnarray} \label{eq_f0}
f(p \leq Q_{\rm s}) = f_0 \quad , \quad f(p>Q_{\rm s}) = 0 \, .
\end{eqnarray}
For the glasma in heavy ion collisions, the phase space is maximally filled: $f_0\sim 1/\alpha_s$ (with $\alpha_{\rm s}\ll 1$). As first emphasized in a recent paper~\cite{Blaizot:2011xf}, such high occupation coherently amplifies scattering and changes usual power counting of scattering rate:
 the resulting collision term from the $2\leftrightarrow 2$ gluon scattering process will scale as $\sim
\alpha_{\rm s}^2 f^2 \sim \hat{o}(1)$  despite smallish $\alpha_{\rm s}$.  This is a natural consequence of the essential Bose enhancement factor $(1+f)$ which would scale as $f$ in the dense regime while scale as $1$ in the dilute regime. It becomes more obvious if one examines the momentum diffusion parameter in Eq.(\ref{qhatel}): $I_a\sim  \hat{o}(1/\alpha_s^2)\, Q_s^3$ and $\hat{q}_{el} \sim \hat{o}(1) \, Q_s^3$, and therefore the time for order one change of typical momentum via scatterings scales as $\tau \sim \hat{o}(1) \, Q_s^{-1}$. With the coupling constant dropping out of the problem,    the system
  behaves as an emergent strongly interacting matter, even though the elementary coupling is small.

A novel finding ~\cite{Blaizot:2011xf,Blaizot:2013lga}, hitherto unrealized, is that a system with such initial condition is highly overpopulated: that is, the gluon occupation number is parametrically large  when compared to a system in thermal equilibrium with the same
  energy density.   To illustrate this point, consider the energy and particle number densities with the initial distribution (\ref{eq_f0}), we have
\begin{eqnarray}\label{initialparam}
 \epsilon_0  = f_0\, \frac{Q_s^4}{  8\pi^2},\qquad n_0=f_0\, \frac{Q_s^3}{6\pi^2},\qquad
n_0 \, \epsilon_0^{-3/4} = f_0^{1/4} \, \frac{2^{5/4}}{3\, \pi^{1/2}} ,
\end{eqnarray}
with $\epsilon_0$ and $n_0$ the initial energy density and number density, respectively. The energy is always conserved during the evolution while the particle number would also be conserved if only elastic scatterings are involved.
The value of the parameter $n \, \epsilon^{-3/4}$ that corresponds to the onset of Bose-Einstein condensation, i.e., to an equilibrium state with vanishing chemical potential,  is obtained by taking for $f(p)$ the  ideal  distribution for massless particles at temperature $T$. One gets then $\epsilon_{SB}=(\pi^2/30)\, T^4$ and $n_{SB}=(\zeta(3)/\pi^2)\, T^3$, so that
\begin{eqnarray}\label{overpopparam}
n \, \epsilon^{-3/4}  |_{SB} = \frac{30^{3/4}\, \zeta(3)}{\pi^{7/2}} \approx 0.28.
\end{eqnarray}
Comparing with $n_0 \, \epsilon_0^{-3/4} $ in Eq.~(\ref{initialparam}), one  sees that when $f_0$ exceeds the value $f_0^c \approx 0.154$, the initial distribution (\ref{eq_f0}) contains too many gluons to be accommodated in an equilibrium Bose-Einstein distribution, i.e. overpopulated: {\it in this case the equilibrium state will have to contain a  Bose-Einstein condensate if there are only elastic scatterings}. It is worth emphasizing that over-occupation does not require necessarily large values of $f_0$, in fact the values just quoted are smaller than unity. It follows therefore that the situation of over-occupation will be met for generic values of $\alpha_s$. For instance, for $\alpha_s\simeq 0.3$, $f_0=1/\alpha_s$ is significantly larger than $f_0^c$ for a wide class of initial conditions (and even more so if the coupling is smaller). One though may also notice that in theories like QCD there are inelastic processes: this removes in principle the possibility of any condensate in the equilibrated state as the inelastic, number changing processes (no matter slow or fast) will eventually remove all the excessive particles. This however leaves open an even more interesting question: starting with overpopulated initial conditions, will the system dynamically evolve and develop a transient Bose-Einstein Condensate?

We therefore see that a Bose system in highly overpopulated regime bears distinctive features that may play key roles in the glasma evolution, including the parametrically enhanced soft elastic scatterings with order one rate and the possibility of  a transient Bose condensate during the course of thermalization. Significant interests and intensive investigations have been triggered recently in understanding such overpopulated regime  with a variety of approaches ~\cite{Kurkela:2011ti,Kurkela:2011ub,Epelbaum:2011pc,Gelis:2011xw,Berges:2012us,Berges:2011sb,Berges:2012ev,Berges:2012iw,Kurkela:2012hp,Schlichting:2012es,Berges:2012ks,Attems:2012js,Iwazaki:2012xi}. There are strong  evidences for Bose condensation  reported for similar overpopulated systems in the classical-statistical lattice simulation of scalar field theory \cite{Epelbaum:2011pc,Gelis:2011xw,Berges:2012us}, with the case for non-Abelian gauge theory still under investigation \cite{Berges:2011sb,Berges:2012ev,Berges:2012iw,Kurkela:2012hp,Schlichting:2012es}. In the rest of this Section, we will first discuss a number of interesting results on the kinetic evolution in such overpopulated systems, including the scaling solutions and  the dynamical onset of kinetic BEC with the purely elastic scattering, and in the last part also discuss the effects of inelastic collisions.

\subsection{The two scales and scaling solutions for elastic scattering}

To qualitatively describe the kinetic evolution, one may  introduce two scales for characterizing a general distribution: a soft scale $\Lambda_{\rm s}$ below which the occupation reaches $f (p<\Lambda_{\rm s}) \sim 1/\alpha_{\rm s}\gg 1$ and a hard cutoff scale $\Lambda$ beyond which the occupation is negligible $f (p>\Lambda)\ll 1$. For the glasma initial distribution in (\ref{eq_f0}),  there is essentially only one scale i.e. the saturation scale $Q_{\rm s}$ which divides the phase space into two regions, one with $f\gg1$ and the other with $f\ll 1$, i.e. with the two scales overlapping $\Lambda_{\rm s} \sim \Lambda \sim Q_{\rm s}$. The thermalization is a process of maximizing the entropy (with the given amount of energy). The entropy density for an arbitrary distribution function is given by $s \sim \int d^3\bp \left[(1+f)\, \ln(1+f)-f \, \ln (f)\right]$: this implies that with the total energy constrained, it is much more beneficial to have as wide as possible a phase space region with $f\sim 1$.  Indeed, for a thermal Bose gas one has the  soft scale $\Lambda_{\rm s}^{th} \sim \alpha_{\rm s} T $ and the hard scale $\Lambda^{th} \sim T$ separated by the coupling $\alpha_{\rm s}$. By this general argument, one shall   expect the separation of the two scales along the thermalization process: from the $\Lambda_s \sim \Lambda$ in the initial glasma toward the $\Lambda^{th}_s \sim \alpha_{\rm s} \Lambda^{th}$ in the thermal situation.

To be more quantitative, one may define the two scales $\Lambda$ and $\Lambda_{\rm s}$ as follows:
\begin{eqnarray}
\Lambda \left( {{\Lambda_{\rm s}} \over {\alpha_{\rm s}}} \right)^2  \equiv I_a  \quad &,& \quad
\Lambda \left({{\Lambda_{\rm s}} \over {\alpha_{\rm s}}} \right) \equiv I_b   \\
{\rm or} \;\;\;\;\;\;\;\Lambda = \frac{I_b^2}{I_a} \quad  &,& \quad \Lambda_s = \alpha_s \frac{I_a}{I_b}
\label{eq_def2}
\end{eqnarray}
With the above definition we indeed have $\Lambda_{\rm s} \sim \Lambda \sim Q_{\rm s}$ for the glasma distribution while $\Lambda_{\rm s}^{th} \sim \alpha_{\rm s} \Lambda^{th} \sim  \alpha_{\rm s}  T$ for thermal distribution. Again one  can see  that with the overpopulated  glasma distribution the collision term ${\cal C}\sim \Lambda_s^2 \Lambda \sim \hat{o}(1)$ in coupling, in contrast to the thermal case with ${\cal C}\sim {\Lambda^{th}_s}^2 \Lambda^{th} \sim \hat{o}(\alpha_s^2)$.

Let us now  discuss possible scaling solution for the evolution of the two scales in the static box case. With the glasma distribution the scattering time from the collision integral on the RHS of the transport equation (\ref{bolel}) scales as $t_{\rm sca} \sim \Lambda / \Lambda_{\rm s}^2$. To find scaling solution for the time evolution of $\Lambda$ and $\Lambda_s$, we use two conditions --- that the energy must be conserved and that the scattering time shall scale with the time itself, i.e.:
\begin{eqnarray} \label{eq_sca}
t_{\rm sca} \sim \frac{\Lambda}{ \Lambda_{\rm s}^2} \sim t   \quad  , \quad \epsilon \sim \frac{\Lambda_{\rm s} \Lambda^3 }{ \alpha_{\rm s} } =  {\rm constant}
\end{eqnarray}
The particle number also must be conserved, albeit with a possible component in the condensate:
$n =  n_g + n_c  \sim (\Lambda_{\rm s} \Lambda^2/ \alpha_{\rm s}) + n_c  =  {\rm constant}$.  The condensate plays a vital role with little contribution to energy while unlimited capacity to accommodate excessive gluons. With these two conditions we thus obtain:
\begin{eqnarray}
\Lambda_{\rm s}  \sim Q_{\rm s} \left( \frac{t_0}{t} \right)^{3/7} \quad , \quad  \Lambda  \sim Q_{\rm s} \left( \frac{t_0}{t} \right)^{-1/7}
\end{eqnarray}
From this solution, the gluon density $n_g$ decreases as $\sim (t_0/t)^{1/7}$, and therefore the condensate density is growing with time, $n_c \sim (Q_{\rm s}^3/\alpha_{\rm s}) [1-(t_0/t)^{1/7}]$. A parametric thermalization time could be identified by the required  $\Lambda_{\rm s} / \Lambda \sim \alpha_{\rm s}$:
\begin{eqnarray}
t_{\rm th} \sim \frac{1}{Q_{\rm s}} \,  \left( \frac{1}{\alpha_{\rm s}} \right)^{7/4}
\end{eqnarray}
At the same time scale the overpopulation parameter $n\epsilon^{-3/4}$ indeed also reduces from the initial value of order $\sim 1/\alpha_s^{1/4}$ to be of the order one.

What would change if one considers the more realistic evolution with boost-invariant  longitudinal expansion?
First of all the conservation laws will be manifest differently:  the total number density will decrease as $n\sim n_0 t_0/t$, while the time-dependence of energy density depends upon the momentum space anisotropy $ \epsilon \sim \epsilon_0 (t_0/t)^{1+\delta}$  for a fixed anisotropy
 $\delta \equiv 	P_L/\epsilon$ (with $P_L$ the longitudinal pressure). Along similar line of analysis as before with the new condition of energy evolution  we obtain the following scaling solution in the expanding case:
\begin{eqnarray}
\Lambda_{\rm s}  \sim Q_{\rm s} \left( t_0/t \right)^{(4+\delta)/7} \, , \,  \Lambda  \sim Q_{\rm s} \left( t_0/t \right)^{(1+2\delta)/7}  \, .
 \end{eqnarray}
With this solution, we see the gluon number density $n_g \sim (Q_{\rm s}^3/\alpha_{\rm s}) (t_0/t)^{(6+5\delta)/7}$, and therefore with any $\delta > 1/5$ the gluon density would drop faster than $\sim t_0/t$ and there will be formation of the condensate, i.e.
$n_c \sim (Q_{\rm s}^3/\alpha_{\rm s}) (t_0/t) [1-(t_0/t)^{(5\delta-1)/7}]$. Similarly a thermalization time scale can be identified through the separation of scales to be:
\begin{eqnarray}
t_{\rm th} \sim \frac{1}{Q_{\rm s}} \,  \left( \frac{1}{\alpha_{\rm s}} \right)^{7/(3-\delta)} \, .
\end{eqnarray}
The possibility of maintaining a fixed anisotropy during the glasma evolution is not obvious but quite plausible due to the large  scattering rate $\sim \Lambda_{\rm s}^2/\Lambda \sim 1/t $ that is capable of competing with the $\sim 1/t$ expansion rate and may reach a dynamical balance. In such a scenario a complete isotropization may never be reached due to longitudinal expansion,  while the system may yet evolve for a long time with a fixed anisotropy between average  longitudinal and transverse momenta.

\subsection{Dynamical onset of Bose-Einstein Condensation}

To more quantitatively understand the kinetic evolution of overpopulated glasma, one needs to numerically solve the transport equation which in the pure elastic case is given by Eqs.(\ref{bolel})(\ref{current}). This has recently been reported in~\cite{Blaizot:2013lga}. The solutions of course depend on the initial conditions. In general,  one expects  two types of solutions: evolution from underpopulated initial conditions leads at late time to a thermal Bose-Einstein distribution function; while evolution starting with overpopulated initial conditions shows a transition, {\it in a finite time}, to a Bose-Einstein condensate. If the initial distribution is specified as the glasma type in (\ref{eq_f0}), then as discussed above, which solution occurs depends on whether $f_0$ is greater or smaller than the critical value $f_0^c$. For simplicity we focus our discussions here mostly on the static box case and will briefly comment on the expanding case at the end.

\begin{figure}[htbp]
	\begin{center}
		\includegraphics[width=4.4cm]{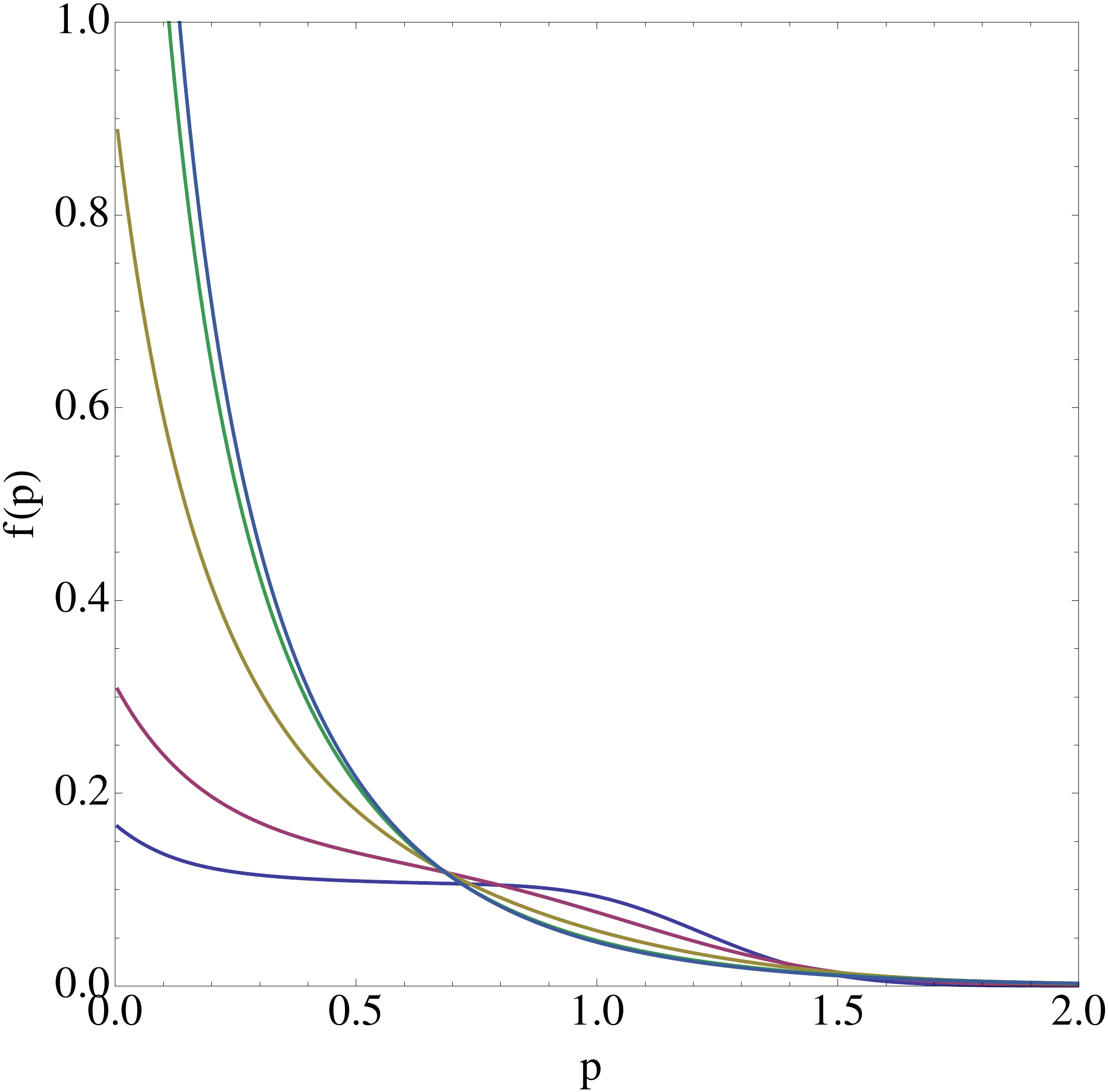} \hspace{0.2cm}
		\includegraphics[width=5cm]{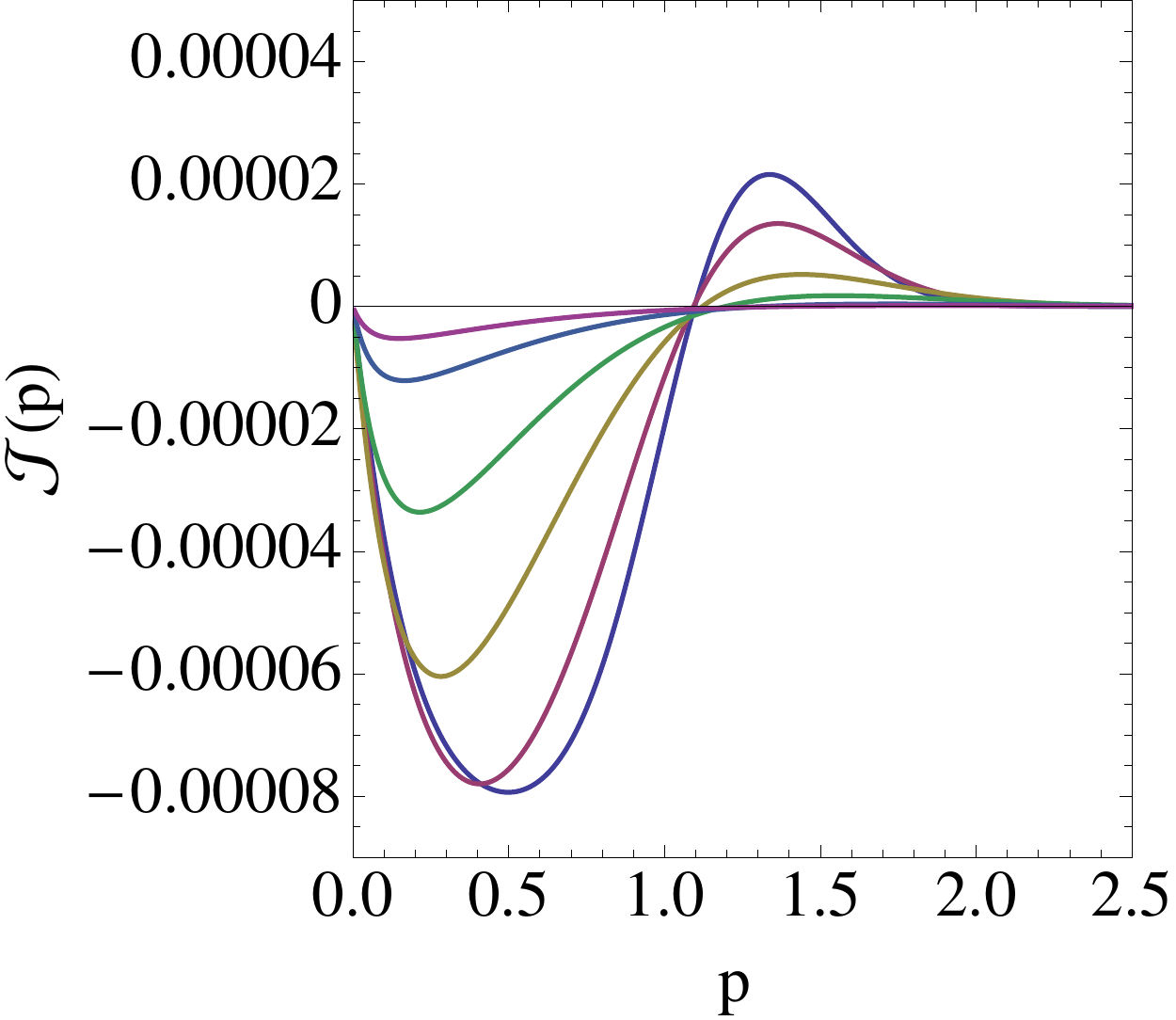}
		\caption{The distribution function $f(p)$ (left) and the current ${\mathcal J}(p)$ (right) for various times, from an early time till the time where thermalization is nearly completed, starting with the underpopulated initial condition $f_0=0.1$.  }
		\label{fig:distr_f01}
	\end{center}
\end{figure}

\begin{figure}[htbp]
	\begin{center}
		\includegraphics[width=5.5cm]{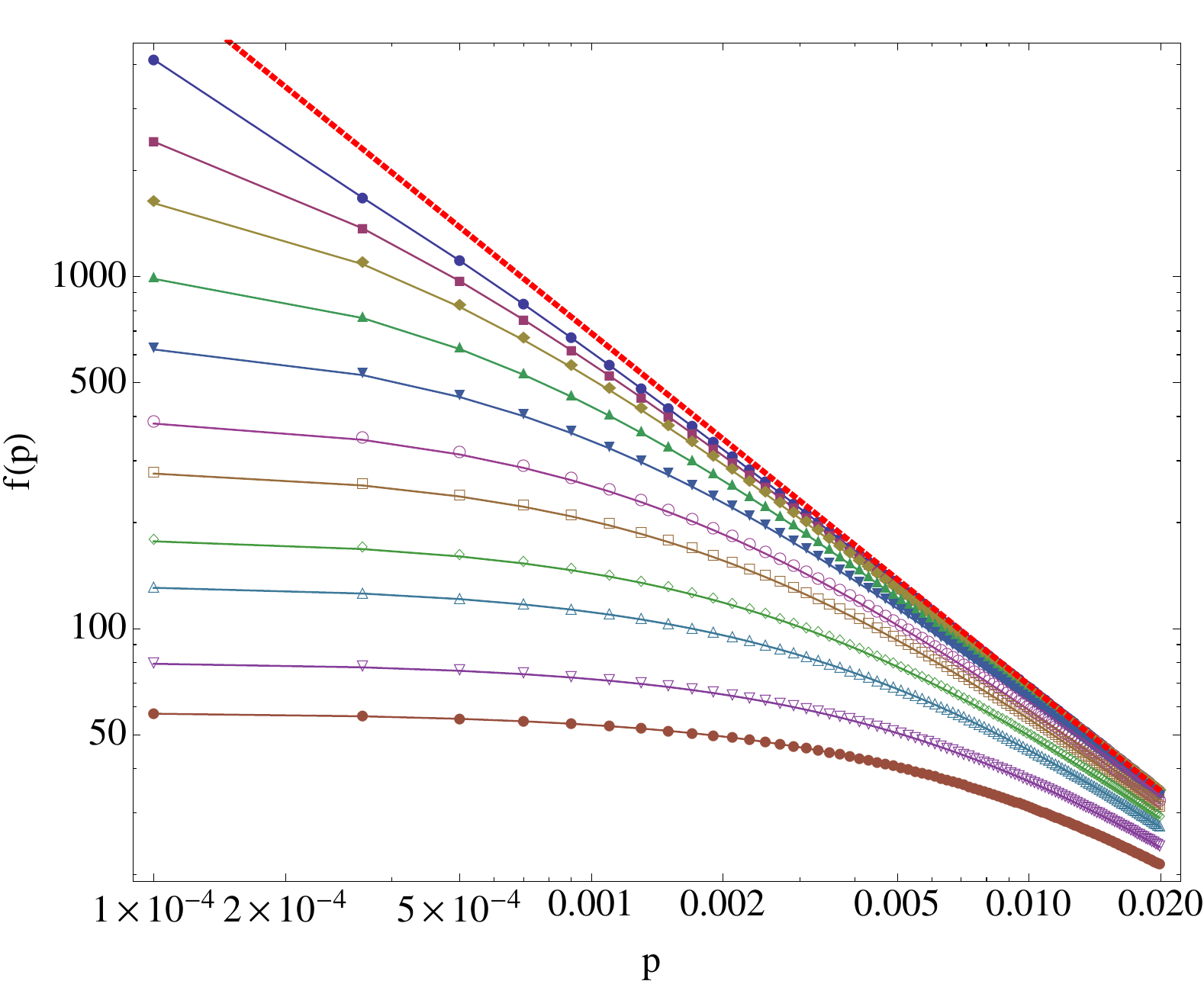} \hspace{0.2cm}
		\includegraphics[width=4.4cm]{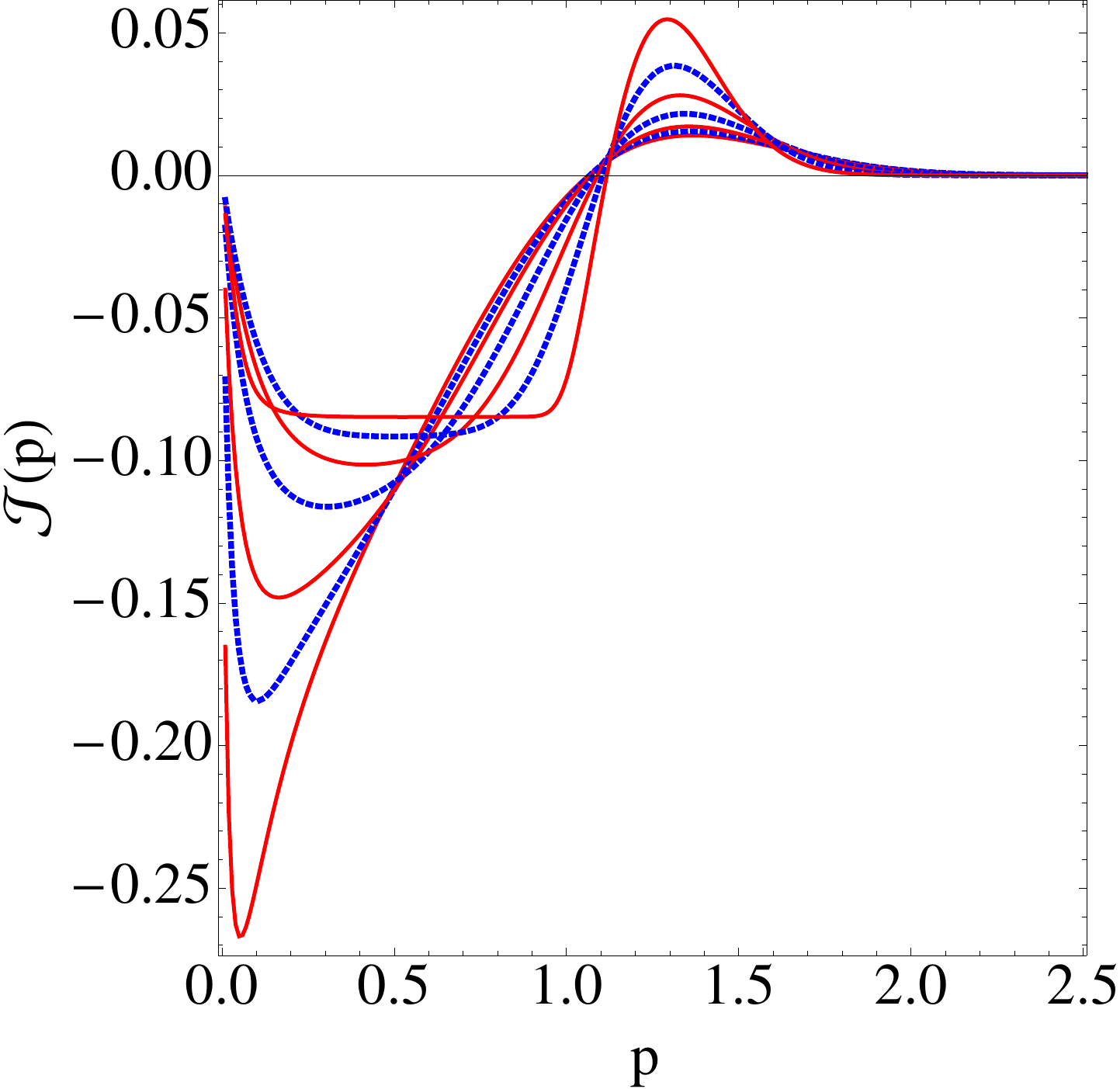}
		\caption{The distribution function $f(p)$ (left) and the current ${\mathcal J}(p)$ (right) for various times, from an early time till the time where thermalization is nearly completed, starting with the overpopulated initial condition $f_0=1$. }
		\label{fig:distr_f1}
	\end{center}
\end{figure}

So how does the thermalization proceed in such a system? Numerical solutions in both the underpopulated and the overpopulated cases suggest two generic features in the kinetic evolution driven by elastic scatterings. First, two cascades in momentum space will quickly develop: a particle cascade toward the IR momentum region that quickly populates the soft momentum modes to high occupation, and a energy cascade toward the UV momentum region that spreads the energy out. The two cascades are of course interrelated as per the particle number and energy conservation.  This can be clearly seen by the plots of the momentum space current (\ref{current}) for the underpopulated (Fig.\ref{fig:distr_f01} right panel) as well as the overpopulated (Fig.\ref{fig:distr_f1} right panel) cases: the negative current at low momenta is the IR cascade and the positive current at high momenta is the UV cascade. It worths emphasizing that the Bose statistical factors play a key role in the strong particle cascade toward IR, amplifying the rapid  growth of the population of the soft modes. As a consequence a high occupation number at IR is quickly achieved, thus with very fast scattering rate, leads to the second interesting feature: an almost instantaneous local ``equilibrium'' form for the distribution near the origin $p\to 0$:
\begin{eqnarray}
f^*(p\to 0)=\frac{1}{{\rm e}^{(p-\mu^*)/T^*}-1},
\end{eqnarray}
Analytically this follows from the requirement that as long as the $f(p\to 0)$ is finite then the current (\ref{current}) has to vanish linearly in $p$ toward the origin. This can be easily seen by integrating the transport equation (\ref{bolel}) in an arbitrarily small sphere around the origin~\cite{Blaizot:2013lga}. The quick emergence of such local IR thermal form has also been numerically verified in both the underpopulated (Fig.\ref{fig:distr_f01} left panel) and the overpopulated (Fig.\ref{fig:distr_f1} left panel) cases. Note that the $T^*$ and $\mu^*$ are only parameters characterizing the small momentum shape of the distribution and not to be confused with a true thermal temperature and chemical potential.

The above picture naturally leads to the next question: how the local IR thermal form eventually evolves into the global thermal form? In the underpopulated case the answer is simple (as explicitly shown by numerical solutions~\cite{Blaizot:2013lga}): the distribution will take time to adjust the whole distribution toward Bose-Einstein distribution (as the proper fixed point of the collision term), with the local parameters $T^*$ and $\mu^*$ approaching the final thermal $T$ and $\mu$ determined by energy and particle number conservation. In the overpopulated case, however, the condensate will need to be formed before the ultimate thermalization. As is well known in the kinetic study of BEC literature~\cite{Semikoz:1994zp,Semikoz:1995rd}, one has to separately describe the evolution prior to the onset of condensation (with the usual transport equation) and the evolution afterwards (with a coupled set of two equations explicitly for condensate and regular distribution).
Of particular significance is to understand dynamically how and when the condensation occurs starting from an overpopulated initial condition. So here let us focus on the  pre-BEC stage, and with the equations (\ref{bolel})(\ref{current}), such question could be answered by numerically solving it till the time of BEC onset.

\begin{figure*}[htbp]
	\begin{center}
		\includegraphics[width=5.cm]{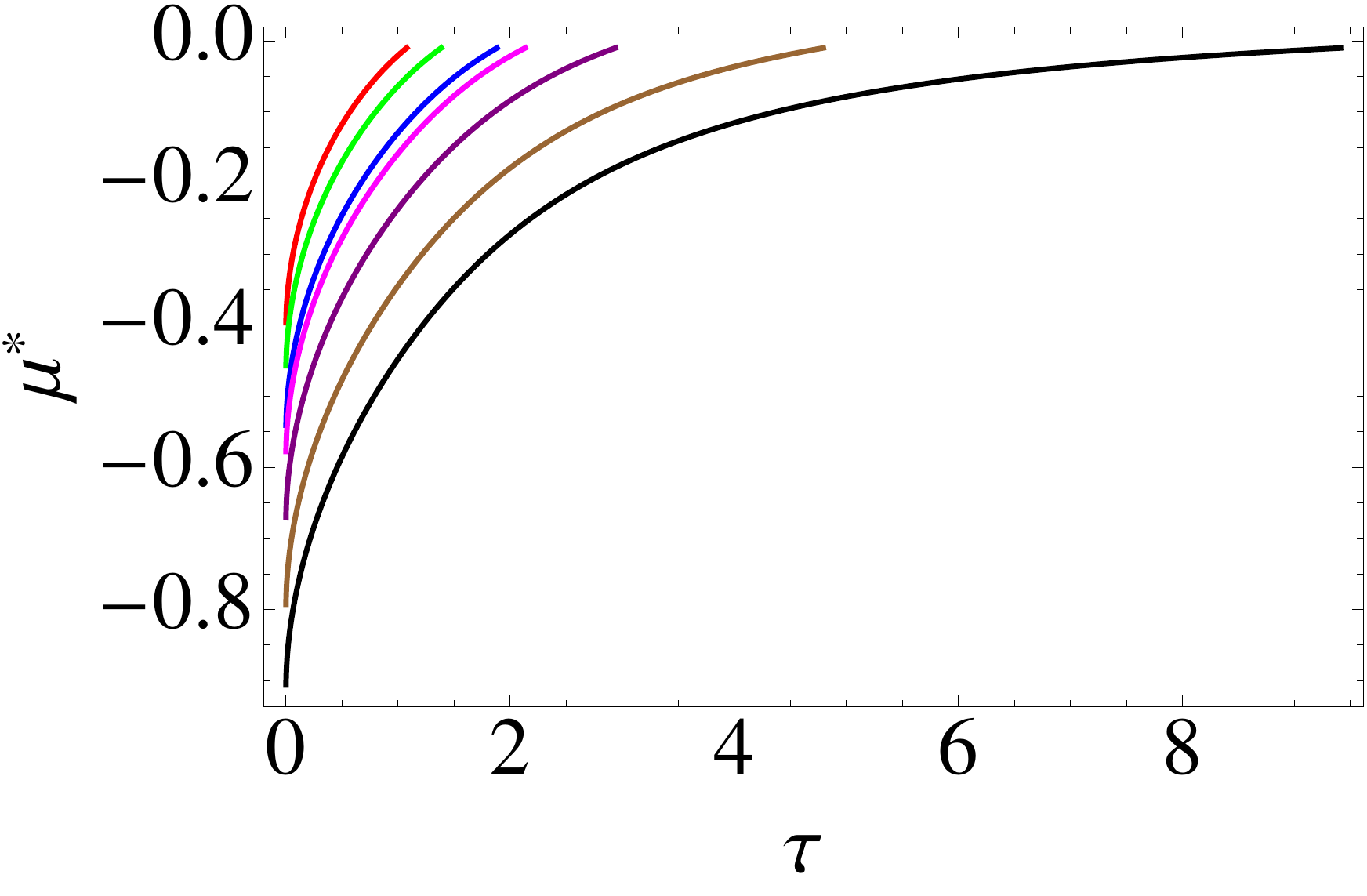}\hspace{0.1cm}
\includegraphics[width=3.3cm]{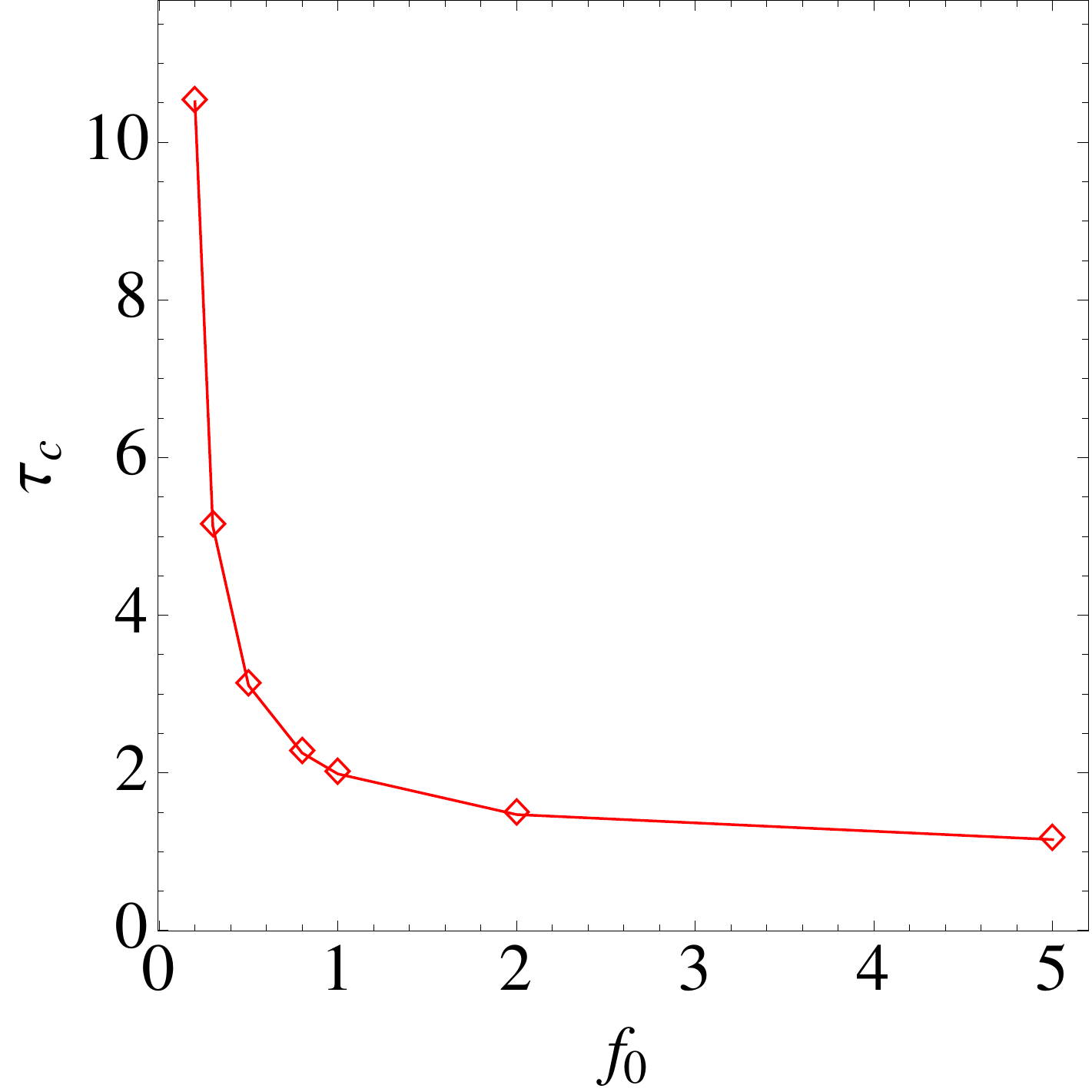}\hspace{0.1cm}
		\includegraphics[width=3.4cm]{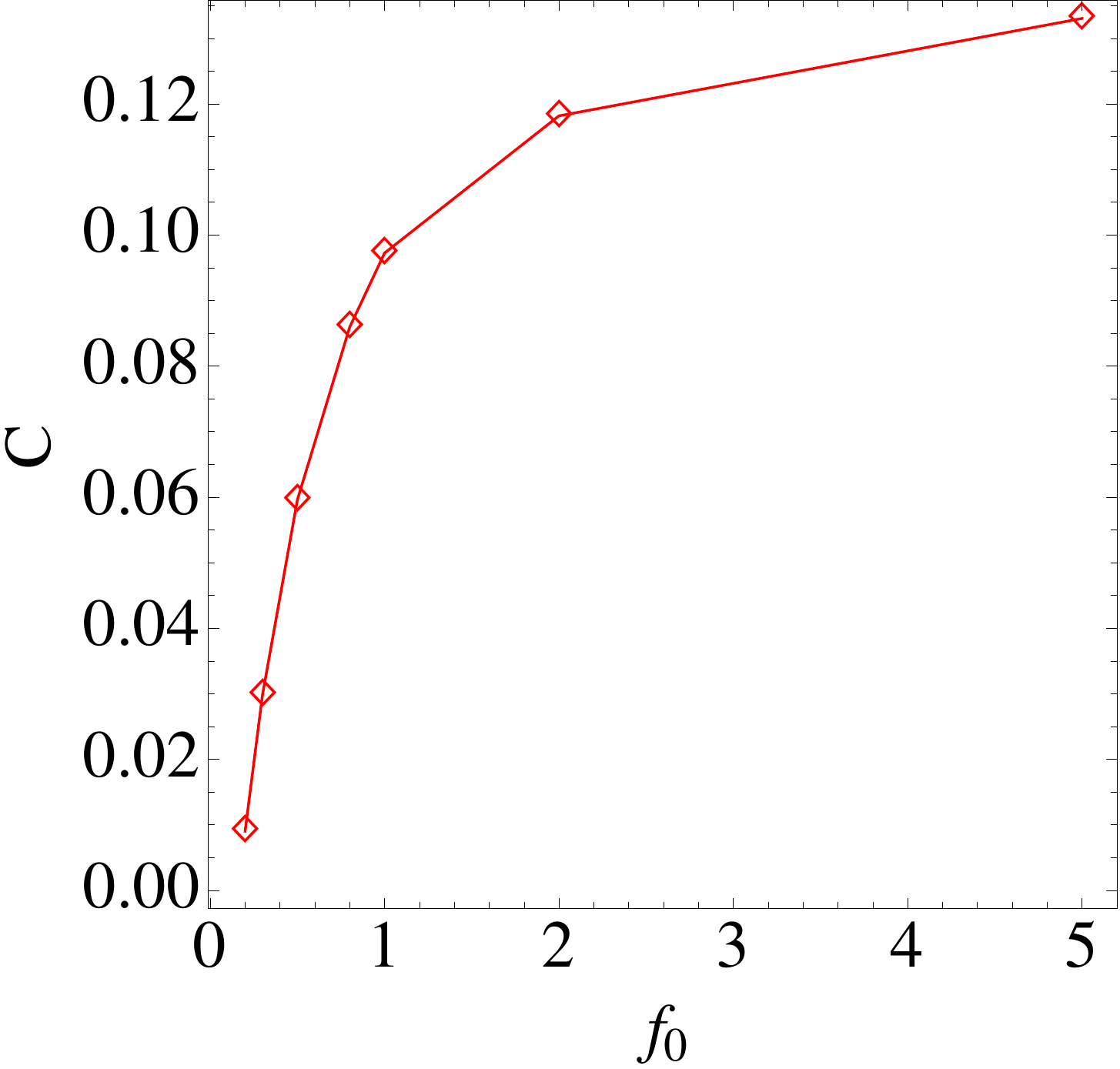}
		\caption{The approach of $\mu^*$ toward zero in a scaling way i.e. $\mu^*\approx C(\tau_c-\tau)$ (left panel) for a variety choice of $f_0$.  One may further extract the value $\tau_c$ (middle panel) at which Bose condensation sets in (left panel) as well as the slope $C$ (right panel) as a function of $f_0$. }
		\label{fig:onset}
	\end{center}
\end{figure*}

In ~\cite{Blaizot:2013lga} this problem has been thoroughly studied  with varied initial conditions and firm evidence has been found that initially overpopulated systems are driven by coherently amplified soft elastic scatterings to reach the onset of Bose-Einstein condensation in a finite time, approaching the onset with a scaling behavior. Different from the underpopulated case, in the overpopulated case the IR cascade persists to drive the local thermal distribution near $p=0$ to increase rapidly in a self-similar form (see Fig.\ref{fig:distr_f1} left panel). The associated negative local ``chemical potential'' is driven to approach zero, i.e. $(-\mu^*) \to 0^+$ (see Fig.\ref{fig:onset} left panel) and ultimately vanishes in a finite time, marking the onset of the condensation. The approaching toward onset is well described by a scaling behavior:
\begin{eqnarray}
|\mu^*|=C(\tau_c-\tau)^\eta
\end{eqnarray}
with a universal exponent $\eta\approx 1 $ for varied values of $f_0>f_0^c$. One may analytically show that the exponent is expected to be unity via similar scaling arguments used in the famous turbulent wave scaling analysis. The onset time $\tau_c$ and the coefficient $C$ is shown in Fig.\ref{fig:onset} (middle and right panels). Such evolution toward onset is robust against different initial distribution shapes, e.g. the same behavior was found with a Guassian initial distribution in the overpopulated regime.

These results, obtained by using  kinetic theory, with a quantum Boltzmann equation in the small angle approximation, have therefore   provided numerical evidence that  a system of gluons with an initial distribution that mimic that expected in heavy ion collisions reaches the onset of Bose-Einstein condensation in a finite time.  The role of  Bose statistical factors in amplifying the rapid  growth of the population of the soft modes is essential.  With these factors properly taken into account, one finds that elastic scattering alone provides an efficient mechanism for populating soft modes, that could be competitive with the radiation mechanism invoked in the scenario of Ref.~\cite{Baier:2000sb}. Ongoing efforts have extended studies of such kinetic evolution toward more general situations, including the effect of longitudinal expansion and possible initial momentum space anisotropy, as well as the effect of finite medium-generated mass. The general link from initial overpopulation to the onset of BEC in a finite time with a scaling behavior appears to be very robust.

There is one particularly important issue, though. It is a prior unclear whether this picture of dynamics BEC onset will be significantly altered, should there be inelastic processes. One may even wonder if such onset (manifested as the development of an infrared singularity in the kinetic evolution) would happen anymore provided any inelastic processes could in principle remove excess particles from overpopulation. To answer this, one needs to study the kinetic evolution including both processes: a first attempt has been done, recently in~\cite{Huang:2013lia}, to be discussed in the next subsection.

\subsection{The effects of inelastic processes}
As we have already seen in Sec.~\ref{kernel23} and Sec.~\ref{lpmeffect}, the peculiar IR enhancement of the QCD makes the effective $1\ra2$ process be comparable to the $2\ra2$ process in the medium. One therefore needs to include both processes in the kinetic evolution. The inclusion of inelastic, number changing process has the immediate consequence that the ultimate thermal equilibrium state can not have any condensate: provided long enough time all excessive gluons can be removed. This however leaves the interesting question: what changes the inelastic collisions bring to the dynamical evolution of the system, and in particular, whether the elastic-driven dynamical onset of condensation from overpopulated initial conditions (as shown in the previous subsection)   would still occur or not.
To answer such question, an explicit evaluation including both elastic and inelastic collisions becomes mandatory. The key issue is the competition between the two kernels: the elastic that drives overpopulated system toward onset of condensation, while the inelastic that tends to reduce the total number density down toward the underpopulation. This problem has recently been addressed in ~\cite{Huang:2013lia}, with surprising finding that is quite different from naive expectations.
\begin{figure}
\begin{center}
\includegraphics[width=5.55cm]{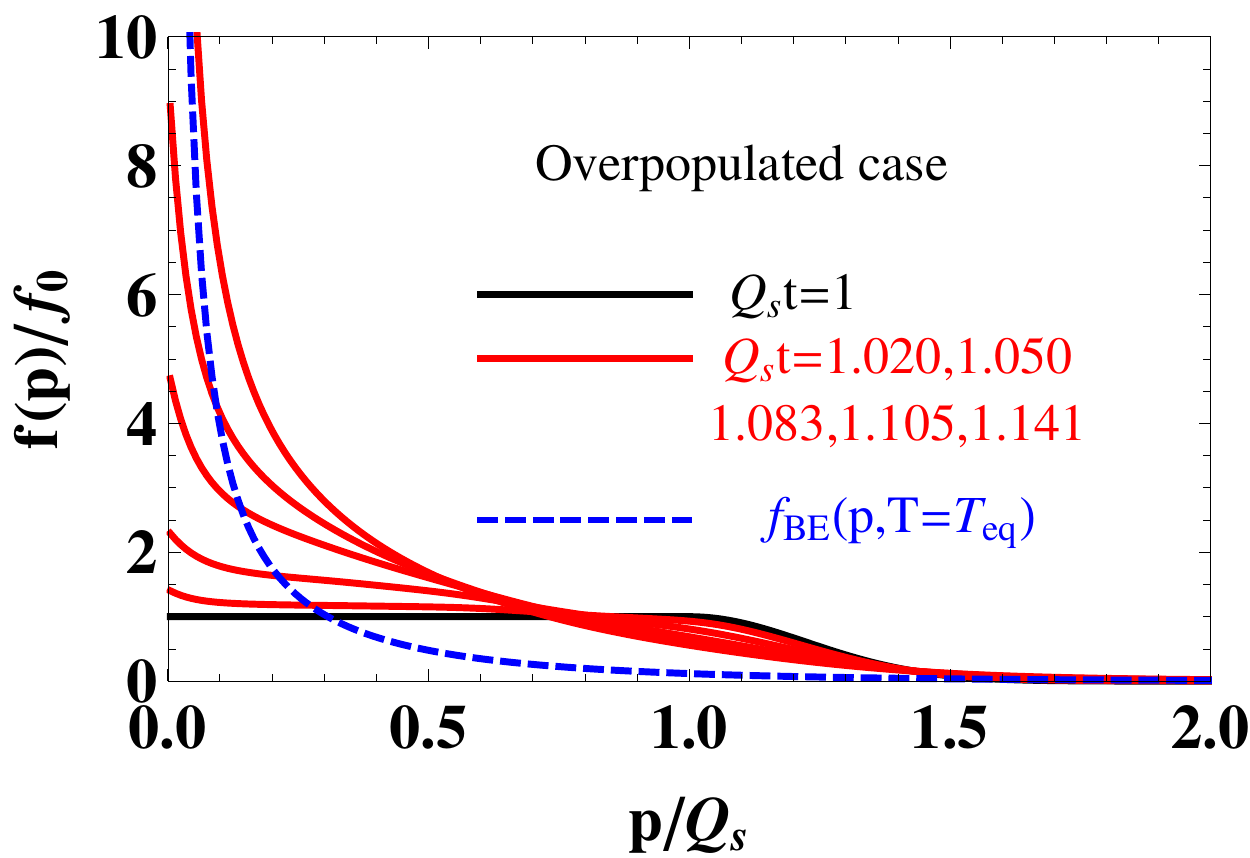}
\includegraphics[width=5.55cm]{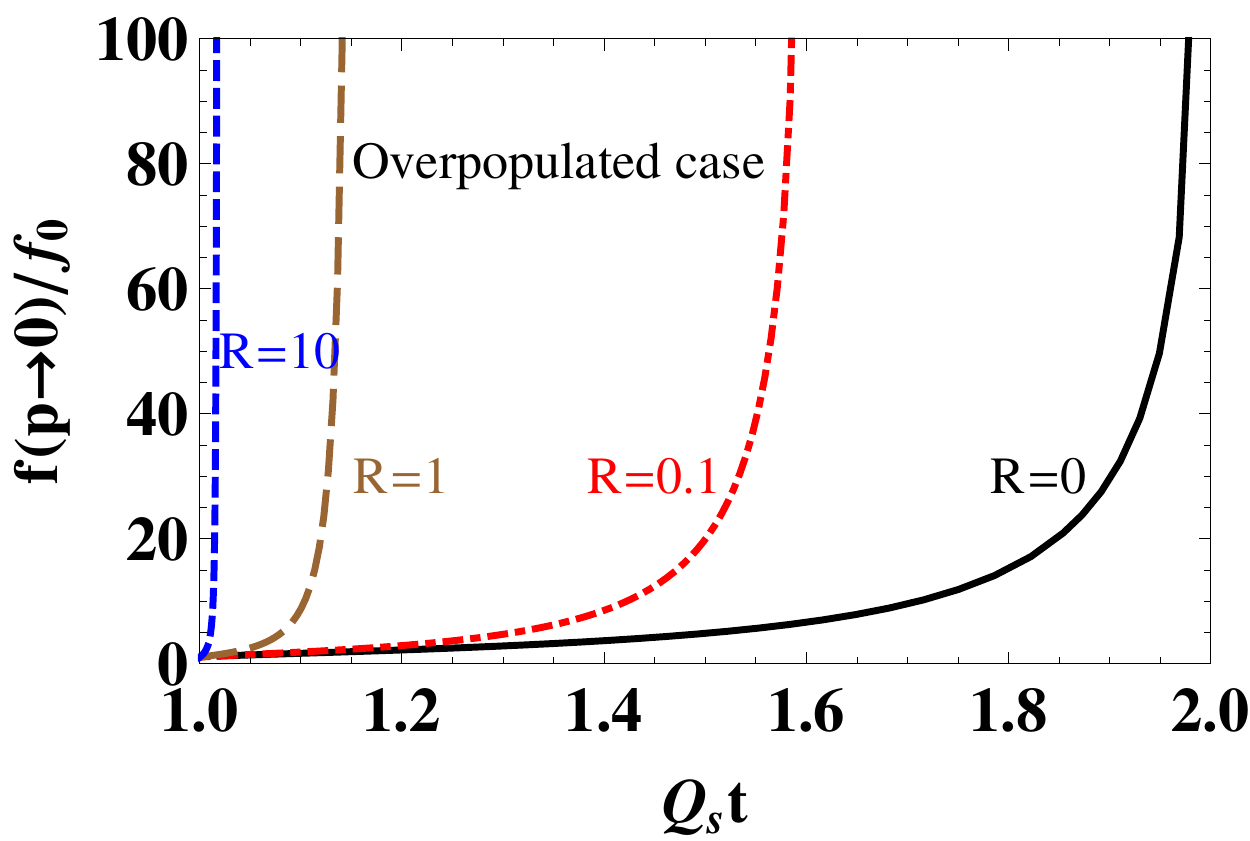}
\caption{(Left)The distribution function $f(p)$ at different time moments. (Right) The occupation near zero momentum as a function
of time for different values of parameter $R$.}
\label{f0_2to3}
\end{center}
\vspace{-0.7cm}
\end{figure}

Starting from overpopulated initial condition in (\ref{eq_f0}) and including both kernels (\ref{bolel})(\ref{eff12}), one can numerically solve the kinetic equation: see \cite{Huang:2013lia} for detailed results and analysis. Let us just highlight the most interesting finding. There is one parameter $R$ that controls the relative strength between the two kernels. As shown in \fig{f0_2to3} (left panel), when the inelastic
processes are turned on, the gluon distribution function at small $p$ region grows very fast (much faster than that with purely elastic process) and quickly becomes a local thermal form $f^*(p)=1/[e^{(p-\m^*)/T^*}-1]$ with the small $p$ part becoming steeper and steeper with time (meaning decreasing $|\m^*|$). This IR evolution proceeds despite that the distribution in the wide range of larger momentum region is still far from equilibrium shape and despite that the overall particle number is indeed dropping.  As a result, the rapid filling of IR modes is enhanced by the inelastic process and the onset of the BEC will occur faster than the purely elastic case. Furthermore as shown in \fig{f0_2to3} (right panel), the stronger (i.e., larger $R$) the inelastic kernel is, the faster the occupation at vanishing momentum will ``explode'' toward the onset of condensation. At first sight this may sound counter-intuitive. To better understand this IR local effect of the inelastic kernel, let us examine the low momentum behavior of   the inelastic kernel:
\begin{eqnarray}
{\cal C}^{\rm eff}_{1\leftrightarrow 2} (p\to 0) \to
R\frac{I_a}{I_b}\ls A_0 f_0(1+f_0)+ A_1 f'_0(1+2f_0) \, p+\hat{O}(p^2)\rs,
\end{eqnarray}
where we have introduced the constants
\begin{eqnarray}
A_0&=&\ln\frac{1}{1-z_c}+\frac{1}{6}\frac{z_c(11z_c^2-27z_c+18)}{(1-z_c)^3},\non
A_1&=&\ln\frac{1}{1-z_c}-\frac{1}{12}\frac{z_c(25z_c^3-88z_c^2+108z_c-48)}{(1-z_c)^4},
\end{eqnarray}
with $z_c$ is an upper cutoff for the integral over $z$.
All these $A$'s are positive for $0<z_c<1$. Clearly for sufficiently small $p$ the leading term in the inelastic kernel $\sim R f_0(1+f_0) A_0$ is {\it always positive} and becomes bigger and bigger with increasing $f_0$ (which is a kind of ``self-amplification''). This leads to extremely rapid growth of the particle number near $p=0$   and the effect becomes stronger with increasing values of $R$, which explains the behavior seen in \fig{f0_2to3}.

Physically this behavior may be understood in two ways. First note that the inelastic kernel has its fixed point to be $1/(e^{p/T}-1)$ which at small $p$ is $\sim 1/p $ so as long as $f(p=0)$ is finite yet the inelastic kernel will try to fill it up toward $1/p$.  Second, this is also related to the quantum effect from Bosonic nature: if all involved particles are from small $p$, then the merging rate is like $\sim f_0^2 (1+f_0)$ while the splitting rate is like $\sim f_0 (1+f_0)^2$ so the splitting ``wins'' due to Bose enhancement for the final state and it increases particle number at small $p$.

\begin{figure}
\begin{center}
\includegraphics[width=7cm]{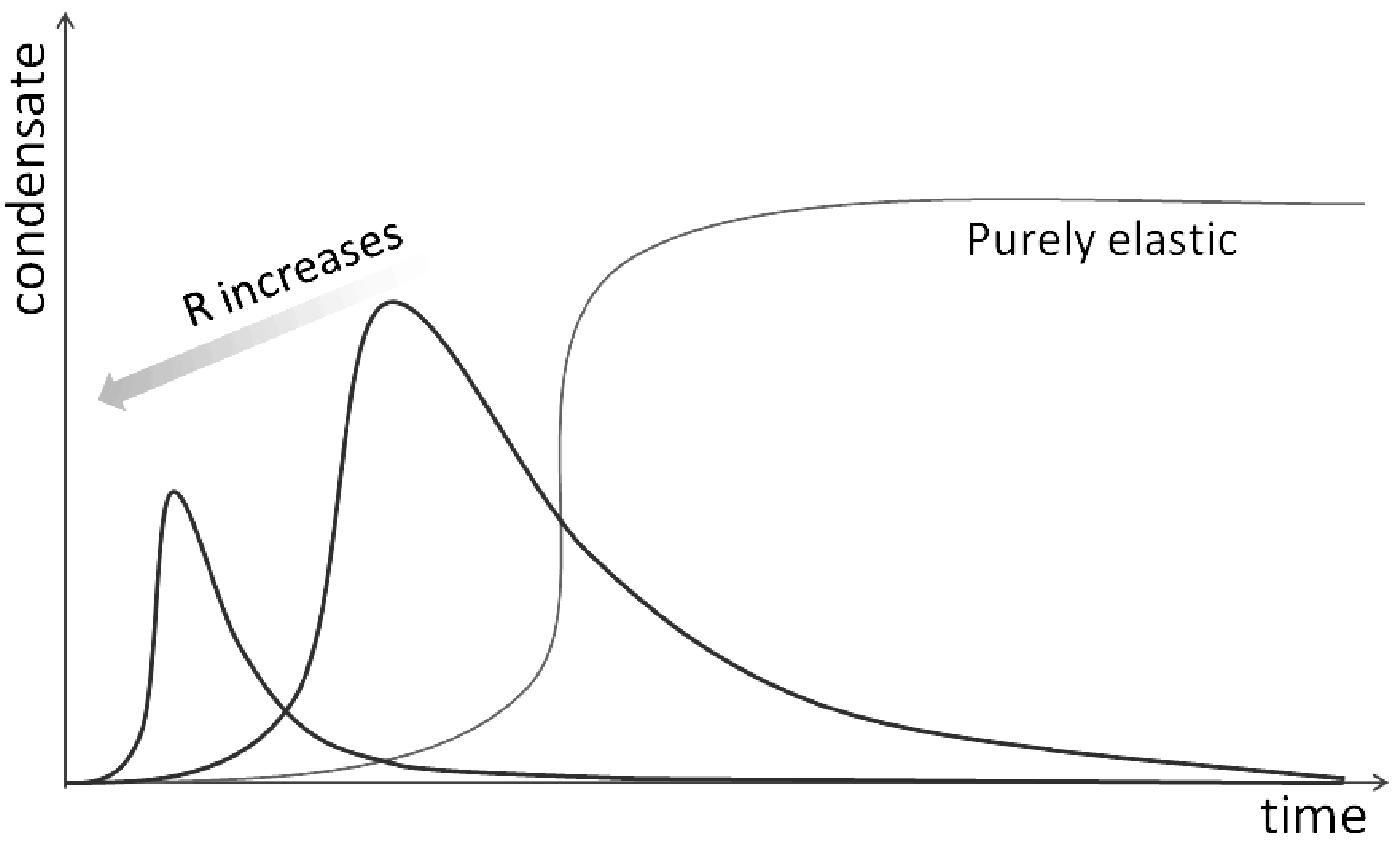}
\caption{Conjectured evolution of the condensate with both elastic and inelastic processes.}
\label{conje}
\end{center}
\vspace{-0.7cm}
\end{figure}

Our finding may sound counter-intuitive at first, as the usual conception would suggest that increasing the strength of the inelastic collisions tends to obstruct more effectively the formation of any condensate. It should however be emphasized that the evolution toward onset that has been studied thus far is not the end of the story. Our analysis addresses the evolution up to the onset of BEC while does not treat the evolution afterwards. As is well known in the BEC literature (see e.g. \cite{Semikoz:1994zp,Semikoz:1995rd}), in order to describe the kinetic evolution of the system with the presence of condensate, a new set of kinetic equations is needed  for an explicit description of the coupled evolution for a condensate plus a regular distribution. Efforts are underway to derive these equations, and so far a kinetic study of the stage after BEC onset for the Glasma system has not been achieved to our best knowledge. However, it appears very plausible that the subsequent evolutions may develop as follows: immediately after onset, the strong IR flux will not cease right away but continue for a while and thus drive the condensate to grow in time; at certain point, the time would be long enough to allow the inelastic processes to decrease the total number density adequately and cause the condensate to decay thus decreasing in time; eventually the inelastic processes will be able to remove all excess gluons and lead to the thermal equilibrium state with neither condensate nor any chemical potential. While the detailed understanding of such dynamic processes can only be achieved  through solving the new set of kinetic equations, one can reasonably expect that with increasing strength of the inelastic processes the whole evolution would be faster. Thus the following overall picture may  likely be the case: with increasing strength, the inelastic processes on one hand catalyze the onset of condensation initially, while on the other hand  eliminate the fully formed condensate faster, thus limiting the time duration for the presence of  condensate to be shorter. A schematic picture of such conjectured full evolution is shown in \fig{conje}, which is in line with the usual conception. It is worth mentioning that recent analysis in \cite{Zhang:2012vi} has shown that the the $2\leftrightarrow3$ inelastic cross section from exact matrix element becomes significantly smaller than that from the Gunion-Bertsch formula, and amounts to $\sim 20\%$ of the $2\leftrightarrow2$ cross section. It therefore seems very plausible that a realistic choice of $R$ value would be rather modest, which may imply a considerable time window for the condensate to be sizable and play an important role for the evolution. A complete investigation of the evolution including the  condensate will be an interesting problem to be pursued in the future.

A final remark concerns the inclusion of quarks (and anti-quarks) into the kinetic evolution of the glasma. So far our discussions have included only gluons, while in reality the quarks and anti-quarks must be there. Even the starting glasma may be overwhelmingly gluonic, quarks and anti-quarks will surely be produced with time via e.g. gluon annihilations into $q\bar q$ pairs. The consequences of adding them are interesting to know. One important change is that the thermal state will have to include the gas of quarks and anti-quarks which change the composition and take a share of the total energy of the system: this will necessarily change the condition for the overpopulation. Another important change is that the individual number conservation for gluons is evaded even without going to higher order complicated multi-gluon scatterings: essentially quarks and gluons can mutually serve as sources via identity-changing processes. On the other hand, one may realize that fermions (subject to Pauli exclusion), unlike bosons, will contribute no more than order $\hat{o}(1)$ to the thermodynamic extensive quantities with each single flavor. Of course one might evade this by dialing large number of flavors, while in reality one has $N_f =3$  which is much smaller than $1/\a_s$ provided $\a_s$ is small. With these general considerations in mind, one may expect that starting with a pure-gluonic, highly-overpopulated initial condition, the gluons may still necessarily reach onset of condensation provided large enough overpopulation,  despite that part of the gluons (about $\sim \hat{o}(N_f)$ ) will be converted into quarks and anti-quarks. Regarding the dynamical evolution of the gluonic sector, one may anticipate a competition between the gluonic elastic scatterings (that drive toward condensation) and the gluon-to-quark conversions (that tend to reduce the gluon overpopulation). A nice and thoroughly quantitative study of this problem has been done very recently by Blaizot, Wu, and Yan~\cite{Blaizot:2014jna} . With a set of kinetic equations that govern the evolution of distributions of both sectors and couple them together, they have found three distinctive behaviors in the solutions from different initial conditions: starting from sufficiently high initial overpopulation, the solution necessarily runs into onset of BEC; starting from initial occupation below certain threshold, the gluon-to-quark conversion is fast enough to completely avoid onset of BEC; while with initial occupation in between the previous two limits, the system reaches a thermal state without gluon condensate but along its evolution runs into a transient stage with gluon condensate. These interesting findings provide further non-trivial evidences for the robustness of the gluon elastic-scattering driven kinetic evolution from overpopulated initial condition toward the dynamical onset of condensation.

\section{Discussions on other kinetic approaches}

While the previous Section has discussed the recent developments emphasizing the role of overpopulation and possible condensation phenomenon, in this Section we also give a brief survey of a number of other interesting studies on the thermalization process in the kinetic framework.

\subsection{The ``bottom-up'' scenario}
A pioneering study in applying the kinetic framework to understand the thermalization in heavy ion collisions was done by Baier, Mueller, Schiff, and Son in~\cite{Baier:2000sb}, where the so-called ``bottom-up" scenario was proposed. In this scenario, one considers a gluon system resulting from the collision between two very large nuclei at extremely high  energy, which is approximately (i) homogeneous in the transverse plane (set as $x$ and $y$ directions), (ii) expanding along the beam direction (set as $z$ axis) in a boost-invariant way, and (iii) having an initial distribution that is highly occupied $f\sim 1/\a_s$ and dominated by ``hard" gluons with momenta $p$ of the order saturation scale $Q_s\gg \L_{\rm QCD}$ and thus $\a_s\ll1$.

The thermalization in this scenario is achieved through three stages. In the first stage, $1\ll Q_s t\ll \a_s^{-3/2}$, the longitudinal expansion dilutes the system and also anisotropizes the distribution according to $p_{x,y}\sim Q_s$, $p_z\sim Q_s/t$ if there were no interactions presented i.e. in free-streaming case. However the small-angle elastic scatterings between hard gluons weaken this anisotropization process via broadening $p_z$ at a rate $dp_z/dt\sim \hat{q}_{\rm el}/p_z$ and as a result the longitudinal momentum is diluted at a slower rate, $p_z\sim Q_s(Q_s t)^{-1/3}$. In this case the hard gluon distribution evolves like $f_h\sim\a_s^{-1}(Q_st)^{-2/3}$. The soft gluons are generated by soft splitting induced by small angle collisions between hard gluons. Once generated, they also ``suffer'' from dilution due to expansion. The combination of these two effects gives the evolution of the soft gluon distribution like $f_s\sim\a_s^{-1}(Q_st)^{-1/3}$ (while overall this whole stage the number density of soft gluons $N_s$ is still much lower than that of hard gluons $N_h$ because they occupy much smaller phase space). At the moment $Q_st\sim\a_s^{-3/2}$ the hard gluon distribution $f_h$ drops from the order $1/\a_s$ initially to the order one by dilution and the system proceeds to the second stage.

In the second stage, the hard sector of the system becomes underpopulated and the hard gluons continue to split into softer ones via inelastic scatterings. The system builds up two scales: one is the hard scale $Q_s$, and the other is the soft scale $k_s\sim\sqrt{\a_s}Q_s$ determined by the screening mass $m_D$ as well as the hard collision rate. The number density of hard gluons continues to drop mainly due to the longitudinal expansion, $N_h\sim (Q_s^3/\a_s) (Q_st)^{-1}$, while the number density of the soft gluons decreases more slowly by virtue of the generation from hard gluon splittings, $N_s\sim\a_s^{1/4}Q_s^3 (Q_st)^{-1/2}$. The distribution of the soft gluons evolves according to $f_s\sim\a_s^{-5/4}(Q_st)^{-1/2}$ which becomes order $1$ at the time $Q_st\sim \a_s^{-5/2}$. After that moment, the system evolves into the third stage.

In the third stage, the soft sector becomes dominant over the hard one while the occupation in both regimes drops below order one. The soft gluons collide frequently and they can isotropize and thermalize fast with small angle scatterings. These then form a ``thermal bath'' with a characteristic temperature $T$ which initially (at the moment $Q_st\sim \a_s^{-5/2}$) is $T\sim k_s$. The hard gluons behave like ``jets" with energy $Q_s$ propagating through this thermal bath and constantly loose their energy into the latter. Therefore the energy is transferred from the hard to soft sector via the LPM-suppressed splitting upon multiple scatterings. The scatterings between the hard gluons themselves are rare due to already low phase space density and can be neglected. Thus the splitting rate is $t_{\rm split}^{-1}\sim \a_s\sqrt{\hat{q}_{\rm el}/k_{\rm split}}$ where $k_{\rm split}$ is the momentum of the emitted gluon and $\hat{q}_{\rm el}\sim\a_s^2 T^3$. The temperature of the soft bath increases until the hard gluons loose all of their energy, which happens when $k_{\rm split}\sim Q_s$. At this point   the system is nearly  thermalized. By equating $t_{\rm split}$ with $t$ and imposing the energy conservation condition $T^4\sim Q_s^4/[\a_s(Q_st)]$, one arrives at a thermalizatoin time $Q_st_{\rm th}\sim \a_s^{-13/5}$ and an equilibrium temperature $T_{\rm eq}\sim\a_s^{2/5}Q_s$.

In the ``bottom-up'' scenario, the overall picture is that soft modes (which can be easily thermalized) will be filed up by hard gluon bremsstrahlung and thermalize first, which then further drains the energy from the hard gluons and make them thermalized, thus the thermalization proceeds from bottom to top in energy scale. We note this scenario differs in two main points from the mostly elastic-driven scenario discussed in the previous Section: first, the elastic scatterings alone are extremely efficient in developing a strong IR flux and provide a mechanism of quickly filling up soft modes, which is absent in the ``bottom-up'' scenario; second, (at least in the static box case) the elastic scatterings also drive a strong UV energy cascade to adjust and thermalize the high momentum tail beyond $Q_s$ scale (a region not discussed in the ``bottom-up'' which may be justified due to expansion) --- how such elastic-driven UV cascade may change by expansion remains to be understood.

\subsection{Instability modified ``bottom-up" approach}
Shortly after the development of the ``bottom-up" scenario, it was realized that there may be complication in the reasoning. This is related to the delicate role of momentum space anisotropy, induced by longitudinal expansion. When the momentum distribution becomes anisotropic, a mechanism completely different from usual scatterings, namely the ``plasma instability'', will occur and play important role. To see that, one may examine the one-loop self-energy tensor $\P^{\m\n}(\o,\bk)$ in a medium with momentum anisotropy,  and in turn the effective propagator of soft gluons is also anisotropic. As it turns out, the dispersion relation obtained from this effective propagator contains branches with negative mass square, or $\im(\o(\bk))>0$ for certain soft momentum region. This implies that such soft modes become unstable and their occupation would grow exponentially. As was first emphasized by Mrowczynski and studied in many later papers~\cite{Mrowczynski:1988dz,Mrowczynski:1993qm,Mrowczynski:2000ed,Romatschke:2003ms,Romatschke:2004jh}, the particularly important instability is the non-Abelian equivalence of the Weibel instability~\cite{Weibel:1959zz}. As a result of such instabilities,  a set of chromo-magnetic modes at scale $k_{\rm inst}$ will exponentially grow to be strong and subsequently diffuse the momenta of hard gluons via Lorentz force to drive the system toward isotropization and thermalizaton. Shortly after the ``bottom-up" scenario, Arnold, Lenaghan, and Moore~\cite{Arnold:2003rq} argued that the plasma instability could be a more efficient mechanism for filling up soft modes and for isotropizing momentum distribution at least for the first stage of the ``bottom-up" scenario and can lead to a faster thermalization at time $Q_st\sim \a_s^{-5/2}$. The roles of plasma instabilities have subsequently been thoroughly analyzed by analytical method~\cite{Bodeker:2005nv}, modified kinetic approaches~\cite{Mueller:2005un,Mueller:2005hj,Mueller:2006up,Arnold:2007cg}, classical field simulations~\cite{Romatschke:2005pm,Romatschke:2005ag,Romatschke:2006nk,Romatschke:2006wg},  hybrid approaches~\cite{Rebhan:2008uj,Rebhan:2009ku,Arnold:2004ih,Rebhan:2004ur,Bodeker:2007fw}, etc.

More recently Kurkela and Moore~\cite{Kurkela:2011ti,Kurkela:2011ub,Kurkela:2012hp,York:2014wja} has carefully analyzed again the roles of plasma instability versus scatterings, particularly in the longitudinally expanding case. An interesting new feature they proposed is that the plasma instability is not only important at the very early stage but may also dominate the thermalization dynamics in all the three stages of the ``bottom-up'' scenario. In the first stage, $1\ll Q_st\ll\a_s^{-8/7}$, the occupancy of both hard and soft gluons decrease and the expansion causes the anisotropy to increase as a function of time, $\lan p_z\ran/\lan p_\perp\ran\sim(Q_st)^{-1/8}(Q_s/p)^{2/3}$. However, the instability causes very fast isotropization for gluons with $p<k_{\rm iso}\sim (Q_st)^{-3/16}Q_s$ and in more infrared region $p<p_{\rm max}\sim(Q_st)^{-1/4}Q_s$ the distribution quickly forms a thermal-like tail which evolves like
$f(p)\sim\a_s^{-1}(Q_st)^{-7/8}(Q_s/p)$. In the second stages, $\a_s^{-8/7}\ll Q_st\ll\a_s^{-12/5}$, the system is highly anisotropic but the hard modes are underpopulated. The hard gluons begin to emit daughter gluons and the anisotropy of hard modes ``propagates''  into the soft region. The plasma instability driven by the anisotropy from these emitted gluons is argued to dominate and control the evolution of $k_{\rm iso}$ as well as $p_{\rm max}$. At the moment $Q_st\sim\a_s^{-56/25}$, $f(p_{\rm max})$ drops to $\sim1$ and the soft sector now forms a nearly-thermal bath with temperature $T\sim p_{\rm max}$. This soft bath does not dominate either energy or scattering at this stage, but it grows to be more and more important and eventually begins to dominate the physics at $Q_st\sim\a_s^{-12/5}$. Then the system enters the third stage $\a_s^{-12/5}\ll Q_st\ll\a_s^{-5/2}$. In this stage, the soft sector (which is weakly anisotropic as a result of   expansion as well as anisotropy passed along from hard gluon splittings) and the resulting  plasma instabilities control the broadening of the hard primary gluons. The instabilities give $\hat{q}_{\rm inst}\sim\a_s^3Q_s^3$ (see ~\cite{Kurkela:2011ti,Kurkela:2011ub}) and thus the splitting scale is given by $k_{\rm split}\sim \a_s^2\hat{q}_{\rm inst} t^2$. Combining this with the energy conservation condition and letting $k_{\rm split}\sim Q_s$ (when the energy cascade stops), one arrives at a thermalization time $Q_st_{\rm th}\sim\a_s^{-5/2}$ and equilibrium temperature scale $T_{\rm eq}\sim\a_s^{3/8}Q_s$. In general, the instabilities would be present due to the inevitable anisotropy brought by the longitudinal expansion and play a role in the IR filing and isotropization. Whether they play a dominant role as compared with various other driving mechanisms, remains to be sorted out.

\subsection{The BAMPS approach}
Quantitative simulations on how the inelastic processes contribute to the thermalization of the gluon system have been carried out by Xu, Greiner and collaborators within the BAMPS (for Boltzmann Approach to MultiParton Scatterings) model~\cite{Xu:2004mz,Xu:2007jv,Xu:2007ns,Xu:2007aa,El:2007vg}. BAMPS is a microscopic transport model based on the kinetic equation \eq{boltzmann}  for on-shell partons with the collision kernel including both the $2\ra2$ elastic and the $2\ra3$ inelastic processes. The main feature of BAMPS is based on the stochastic interpretation of the transition rates which ensure full detailed balance for $2\ra3$ scatterings. BAMPS subdivides space into small cells in which the transition rates are calculated and the gluon distribution function $f(t,\bp,\bx)$ is then extracted~\cite{Xu:2004mz}. In BAMPS framework, the matrix elements \eq{2to2} for elastic $2\ra2$ process and Gunion-Bertsch formula \eq{2to3} for inelastic $2\ra3$ process are used while all the infrared divergences due to small-angle and soft collinear singularities are regularized by introducing the Debye screening mass $m_D$ as an infrared cutoff. This Debye mass is calculated locally in space and is an angle-averaged one so that it is always positive even for anisotropic distribution and no instabilities would be present. The LPM effect is approximately encoded in BAMPS by introducing an infrared cutoff to the transverse momentum $k_\perp$ of the emitted gluon which is determined by requiring the formation time of the emitted gluon to be smaller than the gluon in-medium mean-free path~\cite{Biro:1993qt,Wong:1996ta}.

The simulation results of BAMPS have clearly shown the important contribution of the inelastic processes in filling up the infrared modes and in speeding up the system's evolution toward equilibrium. Different initial conditions, both wounded-nucleons initial condition and CGC-inspired initial condition, have been explored and it has been found that the thermalization time is relatively insensitive to such different choices: for either initial condition, for coupling constant $\a_s\sim0.3$, the gluons in the central region of the collision can be effectively isotropized and kinetically thermalized at a time on the order $t_{\rm eq}\sim 1$ fm. One nontrivial feature found in the BAMPS simulation with the CGC-inspired initial condition is that the soft and hard gluons appear to thermalize at almost the same time scale $Q_st_{\rm eq}\sim [\a_s(\ln\a_s)]^{-2}$ in contrast to the ``bottom-up'' scenario in which the thermalization first occurs in the IR region and proceeds up to the UV region. It would be of great interest to utilize the BAMPS framework and explore the kinetic evolution incorporating full quantum Bose statistics (beyond the classical Boltzmann limit) with overpopulated initial conditions. In particular, it is tempting to see whether a condensation phenomenon may occur or not. For more direct applications to heavy ion collision phenomenology, one may explore within such comprehensive simulation framework the full physical evolution from the initial condition through the thermalization toward the dynamical evolution in the thermal QGP stage, as has been explored in the BAMPS framework as well as in other new transport framework recently developed in e.g.\cite{Ruggieri:2013ova}.

\subsection{Turbulent thermalization and non-thermal fixed point}
As already discussed in Section 2, the classical-statistical lattice simulations provide a first-principle method to explore the thermalization process in the weak coupling and high occupancy limit. Many studies have been done in this framework with a lot of interesting results found~\cite{Berges:2013eia,Berges:2013fga,Mueller:2006up,Berges:2008mr,Khachatryan:2008ys,Fukushima:2011nq,Berges:2012ev,Berges:2012cj,Schlichting:2012es,Fukushima:2013dma}. A very interesting recent finding in ~\cite{Berges:2013eia,Berges:2013fga} is that in the simulation for Yang-Mills gauge theory with very small coupling $\a_s\sim 10^{-4}$ (which allows exploring late time behavior within the classical-statistical framework) and for both the non-expanding and longitudinally expanding cases, the system, after a short transient regime, exhibits universal self-similar scaling solutions with wave turbulence characteristic.

As previously discussed,  the classical-statistical lattice method and the kinetic method have an overlap in the range of validity for occupation $1\ll f\ll 1/\a_s$. With such interesting self-similar solutions found directly from real-time lattice situations~\cite{Berges:2013eia,Berges:2013fga}, it is tempting to see whether such solutions could at least be approximately explained in the more intuitive kinetic picture with microscopic scatterings as scaling solutions to the transport equation. In fact, similar turbulent cascade and self-similar evolutions were found in scalar field theory studies in the context of early universe evolution~\cite{Micha:2002ey,Micha:2004bv} where the appearance of the wave turbulence corresponds to a self-similar non-thermal fixed-point solution of the kinetic equation. Following a similar strategy, the authors of ~\cite{Berges:2013eia,Berges:2013fga} has indeed shown that the self-similar solutions found from simulations can be well approximated by solutions from a kinetic theory of the Fokker-Planck type. To see that, let us for a moment neglect the inelastic processes and examine a kinetic equation of the structure given in \eq{bolel}. In the non-expanding case, it is straightforward to show that the equation allows a scaling solution of the general form $f(t,\bp)=(Q_st)^\a f_S((Q_st)^\b\bp)$ provided two conditions:  the stationary function $f_S(\bp)$ satisfying $\a f_S(\bp)+\b\bp\cdot\nabla_\bp f_S(\bp)+Q_s^{-1}\nabla_\bp\cdot{\bf\cal J}[f_S(\bp)]=0$, and the scaling parameters satisfying a relation $\a-1=3\a-\b$. Furthermore, the energy conservation implies an additional relation, $\a=4\b$. Combining these two relations, one obtains $\a=-4/7$ and $\b=-1/7$, which turn out to nicely reproduce the exponents extracted from their classical-statistical simulations.

In the expanding case, the analysis is less straightforward due to the expansion and the competing effect of scatterings in kinetic theory. Since such self-similar solution emerges at relatively later time in the evolution, the system becomes much diluter and the effect of scatterings may be plausibly approximated by pure elastic   momentum broadening in the $z$ (beam) direction, ${\cal C}[f]=\hat{q}_{\rm el}\pt^2_{p_z} f$ with $\hat{q}_{\rm el}$ given by \eq{qhatel}. Such a highly simplified Boltzmann equation $[\pt_t-(p_z/t)\pt_{p_z}]f={\cal C}[f]$ does allow a scaling solution of the form $f(t,\bp_\perp,p_z)=(Q_st)^\a f_S((Q_st)^\b\bp_\perp,(Q_st)^\g p_z)$ with the scaling parameters satisfying $2\a-2\b+\g+1=0$.  Furthermore when the system becomes diluter at late time, its evolution may approach free-streaming case, with the energy and particle number densities both dropping as $\sim 1/t$.  Under such assumptions, one arrives at the solution with exponents $\a=-2/3,\b=0,\g=1/3$. As the authors~\cite{Berges:2013eia,Berges:2013fga} have shown,   these scaling parameters obtained in the kinetic equation are in surprisingly excellent agreement with the self-similar behavior seen in their classical-statistical simulations. It has also been numerically checked that with the given coupling constant regime such self-similar evolution   is insensitive to the initial condition and at very late time it approaches toward the original ``bottom-up" scenario. In general, the appearance of the non-thermal fixed point will delay the thermalization of the system toward the true thermal fixed point. It is worth commenting on the roles of the very small value of coupling used in these studies: technically it allows the classical field approach to be a better controlled approximation with much longer evolution time; physically it opens a very wide window for the occupation (in the kinetic picture) in between the saturated limit $f\sim 1/\a_s$ and the quantum limit $f\sim 1$, likely maximizing the manifestation of nonlinear effects such as turbulent cascade. These findings are extremely interesting, and leave open a number of questions to be explored further, in particular, what change may happen to this scenario when one gradually moves toward the coupling constant regime $\a_s\sim 10^{-1}$ that may be more directly relevant to the glasma in heavy ion collisions.

\section{Summary}

In summary, we have given a brief review of the thermalization problem in heavy ion collisions, with emphasis on recent progress in understanding the kinetic evolution of the glasma. A short discussion has been given on the general context of the thermalization problem and a number of approaches other than the kinetic one. We have then provide an elementary introduction on the transport framework to be used for describing the pre-equilibrium evolution, including both the elastic and inelastic collisions. Recent interesting developments on the kinetic evolution in the {\it overpopulated regime}, as in the case for the glasma, and the possibility of dynamical Bose-Einstein Condensation in such system, have been discussed in details. Finally a number of other approaches within the kinetic framework have been surveyed.

Though there have been a lot of interesting developments and some nontrivial ideas in the last few years, it may be fair to say that we are still far from a detailed understanding of the kinetic evolution for the pre-equilibrium stage in heavy ion collisions. For the kinetic approach in the overpopulated regime, a number of pressing issues need to be understood, including the co-evolution of the condensate and regular distribution (after the onset of condensate), the far-from-equilibrium medium effects (e.g. the dressing of internal/external gluons involved in a scattering), the roles of higher order processes, etc. It is also of great importance to further explore the relation (the overlap in their applicability and their complementarity) between the kinetic description and the classical field description, in particular how certain behavior (e.g. condensation and turbulent scaling) observed in one description would be manifested in the other description. Toward more phenomenological end, it is crucial to implement and investigate the effects of longitudinal expansion as well as the roles of anisotropy (both that from initial condition and that dynamically generated from expansion). It is also highly interesting to study the kinetic evolution with more realistic transverse distributions e.g. by introducing transverse-position dependent initial conditions (via saturation scale) which would allow determining possible early transverse flow generation.  The condensate, if formed, would play nontrivial roles in many aspects of phenomenology from pA to AA collisions, as explored by a number recent studies along this direction~\cite{Floerchinger:2013kca,Dumitru:2014nka,Dumitru:2013tja,Majumder:2014qpa}, and there are certainly many more possibilities to be fully investigated.

Let us end with a discussion on the interesting evolution of the very conception of the problem itself. The initially perceived ``thermalization'' problem , as the name suggests it, has the implicit picture of two distinctive stages:  a pre-thermal stage with the system evolving to a (relatively) complete local thermalization (and of course isotropization) and a thermal stage which then expands in a nearly ideal hydrodynamic fashion, with the switch between the two stages at rather early time $\sim 1$ fm/c. This was largely motivated by the phenomenological success of the ideal hydrodynamic simulations at the early RHIC era (see the nice discussion in the recent review article ~\cite{Strickland:2013uga}), along with the conventional wisdom that the applicability of hydrodynamics requires local thermal equilibration. However there has been no direct evidence for full thermalization (and not even for isotropization). In fact, the later developments of viscous hydrodynamic studies have demonstrated that even with extremely small dissipation (i.e. $\eta/s$ close to the conjectured lower bound) the stress tensor bears sizable anisotropy between longitudinal/transverse pressures over several fm/c time window ~\cite{Strickland:2013uga}. There have also been interesting works from both strong coupling approach (via the holographic models)~\cite{Heller:2011ju,Heller:2012je} and weak coupling approach (via real time lattice simulations)~\cite{Gelis:2013rba} that show the emergence of (viscous)hydrodynamic behaviors without reaching either isotropization or full thermalization. To add to the complications, most recent experimental measurements of high multiplicity pPb collisions at LHC and dAu collisions at RHIC show very interesting patterns in the soft particle productions and correlations, which seem to be accountable by collective expansions akin to viscous hydrodynamic simulations applied to such systems much smaller in size and much shorter-lived in time as compared with the bulk matter in AA collisions~\cite{Bozek:2011if}  (noting though whether this is indeed so is still under intensive debate~\cite{Bzdak:2013zma,Dusling:2013oia,Zhang:2013oca,Shuryak:2013ke} and subject to conclusion in the future).  All these may call for a change in our very identification of the ``thermalization'' problem, splitting into two closely related but clearly different aspects: theoretically how and when a full thermalization is achieved in a quark-gluon system starting with initial conditions close to that in the heavy ion collisions; phenomenologically, how and when an apparent hydrodynamic behavior emerges from the pertinent initial conditions and how far one can push the limit (e.g. in the system size, in the anisotropy, in the dissipation, in the microscopic coupling, etc) for the system to stay amenable to a collective expansion. It will require significant future efforts to fully explore both of these issues and make progress in understanding the ``thermalization'' problem.

\section*{Acknowledgements}

The authors are grateful to J. Berges, J.-P. Blaizot, F. Gelis, L. McLerran, R. Venugopalan, Q. Wang, B. Wu, Z. Xu, and P. Zhuang for discussions and communications. XGH is  supported by Fudan University grants EZH1512519 and EZH1512600. The research of JL is supported by the National Science Foundation under Grant No. PHY-1352368. JL thanks the RIKEN BNL Research Center for partial support. JL is also grateful to the Yukawa Institute for Theoretical Physics, Kyoto University, where this work was partly completed during the YITP-T-13-05 on ``New Frontiers in QCD''.

\end{document}